\documentclass[aps,pre,reprint,nofootinbib,superscriptaddress]{revtex4-2}

\usepackage[T1]{fontenc}
\usepackage[utf8]{inputenc}
\usepackage{amsmath}
\usepackage{amssymb}
\usepackage{newtxtext}
\usepackage{float}
\usepackage{graphicx}
\usepackage{hyperref}
\hypersetup{colorlinks=true,linkcolor=blue,urlcolor=blue,citecolor=blue,pdfstartview=FitH}
\begin{document}

\title{Classical description of the parameter space geometry in the Dicke \\and Lipkin-Meshkov-Glick models}

\author{Diego Gonzalez}
\email{dgonzalez@fis.cinvestav.mx}

\affiliation{Departamento de F\'isica, Cinvestav, Avenida Instituto Polit\'ecnico Nacional 2508, San Pedro Zacatenco, 07360, Gustavo A. Madero, Ciudad de M\'exico, Mexico}
\affiliation{Departamento de F\'isica de Altas Energ\'ias, Instituto de Ciencias Nucleares, Universidad Nacional Aut\'onoma de M\'exico, \\Apartado Postal 70-543, Ciudad de M\'exico, 04510, Mexico}

\author{Daniel Guti\'errez-Ruiz}
\email{daniel.gutierrez@correo.nucleares.unam.mx}

\author{J. David Vergara}
\email{vergara@nucleares.unam.mx}

\affiliation{Departamento de F\'isica de Altas Energ\'ias, Instituto de Ciencias Nucleares, Universidad Nacional Aut\'onoma de M\'exico, \\Apartado Postal 70-543, Ciudad de M\'exico, 04510, Mexico}

\date{\today}

\begin{abstract}
	
We study the classical analog of the quantum metric tensor and its scalar curvature for two well-known quantum physics models. First, we analyze the geometry of the parameter space for the Dicke model with the aid of the classical and quantum metrics and find that, in the thermodynamic limit, they have the same divergent behavior near the quantum phase transition, as opposed to their corresponding scalar curvatures which  are not divergent there. On the contrary, under resonance conditions, both scalar curvatures exhibit a divergence at the critical point. Second, we present the classical and quantum metrics for the Lipkin-Meshkov-Glick model in the thermodynamic limit and find a perfect agreement between them. We also show that the scalar curvature is only defined on one of the system's phases and that it approaches a negative constant value. Finally, we carry out a numerical analysis for the system's finite sizes, which clearly shows the precursors of the quantum phase transition in the metric and its scalar curvature and allows their characterization as functions of the parameters and of the system's size.

\end{abstract}

\maketitle

\section{Introduction\label{sec:Introduction}}

Geometry has found many applications in various areas of science, especially in physics~\cite{Nakahara}. An essential element of geometry is the metric tensor, which contains the relevant information to measure distances in an underlying space. Provost and Vallee~\cite{Provost} endowed the Hilbert space with a metric that is invariant under parameter-dependent phase transformations (gauge invariance), now known as the quantum metric tensor (QMT). It measures the distance in the parameter space of a system and, therefore, encodes the information of how close two quantum states are. The QMT can be obtained from a second-order expansion of the fidelity, which has been used extensively in the study of quantum phase transitions (QPTs)~\cite{GuReview,Damski2011,Paunkovic2018}. A QPT is characterized by a change in the ground state function's analytic properties and separates the system into two different regions in parameter space~\cite{Sachdev}. The pioneering works~\cite{Zanardi2006,Zanardi2007Information} showed the relevance of the QMT and its scalar curvature to study QPTs. Further features of the QMT, such as its scaling properties and critical exponents, have been examined~\cite{Zanardi2007Scaling,SarkarCritical2014,SarkaRenormalization2015}. The geometrical aspects of the QMT have also been looked into, like its geodesics and its scalar curvature, as well as their relation to topology~\cite{SarkarGeodesics2012,KolodrubetzPRB,YQMa2014,KolodrubetzRep,Panahiyan2020}. Interestingly, the relation of the QMT to complexity in the context of quantum computing~\cite{Mancini2018} has also been analyzed. Aside from its theoretical relevance, the QMT can be measured experimentally, providing a direct link to condensed matter systems~\cite{Ozawa2018-1,*Ozawa2018-2,*Ozawa2019}. A detailed account of the geometry of QPTs can be found in Ref.~\cite{Carollo2020}. For an application of geometrical concepts in the study of QPTs in a different context, see Ref.~\cite{Chang2018}.

The QMT possesses a classical analog first introduced in Ref.~\cite{GonzalezPRE}. This classical analog, from now on, called the classical metric, measures the distance in the parameter space between two points in phase space infinitesimally separated; it is defined for classical integrable systems and relies on the adiabatic theorem, introducing a torus average over the angle variables to obtain its components. In Ref.~\cite{GonzalezAnnalen}, it was proved that the classical metric results from a semiclassical approximation of the QMT under the time-dependent Lagrangian approach introduced in~\cite{Alvarez2017}. The classical metric possesses the same properties as its quantum counterpart: it is positive semidefinite, gauge invariant, and it transforms as a rank-two covariant tensor. Given a quantum Hamiltonian, one may try to study its semiclassical version using coherent states. In this case, the QMT can be computed as in Ref.~\cite{Provost}, yielding a parameter space with a flat, spherical, or hyperbolic geometry. Nevertheless, the dependence of the coherent states' coordinates on the Hamiltonian parameters might constitute a noninvertible mapping, which then results in a QMT whose components are zero in one or both phases of the system. Therefore, this semiclassical version of the QMT is useless to characterize the geometry of the parameter space~\cite{Vergara2021PRB}. On the other hand, the classical metric shows its relevance emerging as a tool that, through purely classical functions and a classical torus average, provides a result consistent with the quantum description in many cases. We must mention, however, that some differences between the classical and quantum metrics may appear essentially due to operator-ordering ambiguities, which may result in (i) anomalies that contribute with additional terms~\cite{GonzalezAnnalen, RevModPhys.90.015001},  (ii) differences coming from the fact that there might be distinct quantizations for a given classical Hamiltonian.

In~\cite{GonzalezPRE,GonzalezAnnalen}, some examples that illustrate the use of the classical metric were laid out, and it was shown that the classical metric contains the same or almost the same information about the parameter space as the QMT; however, the need to delve into more geometrical details was manifest. In this sense, it is worth considering the scalar curvature, which is a local invariant of a metric space that quantifies the deviation of this space from being Euclidean~\cite{Villani2008}. We can mention three essential features of this quantity. First, the value of the scalar curvature at a given point is independent of the parameter space's coordinates, which means that it is a geometric invariant quantity. Thus, it is  helpful to detect whether a singularity is real or it is merely an effect of the parameter space's coordinates that are being used~\cite{Plebanski}. This fact contrasts with the classical and quantum metrics, which depend on the choice of parameters and may contain removable singularities. Second, not all the systems have a Berry curvature and, consequently, a Berry phase. However, even in these cases, there might be a nonvanishing scalar curvature that will help characterize the underlying geometry.  Third, the scalar curvature can also be used to obtain global information of the manifold, like the Euler characteristic, which is a topological number used to investigate the nature of the singularities associated with QPTs and provides a characterization of each quantum phase~\cite{KolodrubetzPRB}. Naturally, the scalar curvatures coming from the classical and quantum metric tensors for a given system can be computed and compared to see whether the classical treatment yields the same information as its quantum counterpart.

A first approach to test the classical methods is to examine quantum many-body systems with a QPT in the thermodynamic limit. Of the variety of models that can appear, two of them have been widely used due to their rich parameter space structure and the effectiveness of a classical treatment: the Dicke model and the Lipkin-Meshkov-Glick (LMG) model. The Dicke model~\cite{Dicke} consists of two-level atoms interacting with one mode of a bosonic field inside a cavity. Its parameter space geometry has been studied in Ref.~\cite{Sarkar2012}, where the thermodynamic limit was considered. On the other hand, the LMG model~\cite{Lipkin1,*Lipkin2,*Lipkin3} describes the interaction of spin-half particles interacting with each other and with an external magnetic field. A brief account of the QMT in this model can be found in Refs.~\cite{GuPRE,Sarkar2012}. However, further geometrical analysis is lacking.

Our goal in this paper is twofold: first, to test the limits of the classical metric for the Dicke and LMG models, and see how it departs from the quantum description; second, to study the scalar curvature of the classical and quantum metrics and its behavior near the QPT to extract valuable information of the critical region. We recall that, in a two-dimensional (parameter) space, which is the case considered here, the scalar curvature  characterizes the geometrical structure of the surface. We will find the Dicke model's classical and quantum metrics in the thermodynamic limit under the truncated Holstein-Primakoff approximation, where the system becomes integrable.  Although the metrics are not equal, they turn out to be singular at the QPT and diverge in the same manner, whereas their scalar curvatures are not divergent in the critical region. When the resonance condition is considered, both metrics and their scalars show divergence at the QPT. In the case of the LMG model in the thermodynamic limit, the classical and quantum metrics are the equal up to a quantization rule for the action variables, and the scalar curvature diverges at the QPT. In this paper, the classical metric and its scalar curvature are obtained for the Dicke and LMG models, and it is remarkable that they are able to give the same information as their quantum counterparts at the QPT. We also perform a detailed numerical study of the QMT and its scalar curvature for finite sizes of the LMG model, which extends the results of~\cite{GuPRE} where only the fidelity susceptibility (i.e., a component of QMT) is analyzed. In this case, we will see the QPT precursors in the peaks of the metric components and the scalar curvature, and an extrapolation of the results to the thermodynamic limit will give a clue as to whether the singularities predicted by the classical metric are genuine or are just a consequence of the choice of parameters. This will shed light on previous works where the scalar curvature's behavior for some models in the critical region was investigated~\cite{Zanardi2007Information,Sarkar2012}.

The structure of the paper is as follows. In Sec.~\ref{sec:Geometry}, we begin by reviewing the essential geometric elements that will help us describe the parameter space of a system; we present the QMT following the approach of Ref.~\cite{Alvarez2017}, introduce its classical analog~\cite{GonzalezAnnalen}, and discuss some features of the scalar curvature in two dimensions. In Sec.~\ref{sec:Dicke}, we analyze the classical and quantum metrics for the Dicke model in the thermodynamic limit and compare their corresponding scalar curvatures for the nonresonant and resonant cases. In Sec.~\ref{sec:Lipkin}, we consider the LMG model and we compute the associated classical and quantum metrics in the thermodynamic limit as well as their scalar curvatures. Then, for finite sizes of the LMG model, we obtain numerically the QMT and its scalar curvature. We also analyze the peaks of these numerical quantities as well as the slope of the scalar curvature at the QPT, and deduce their behavior in the thermodynamic limit. Finally, in Sec.~\ref{sec:Conclusions}, we present the conclusions and propose some aspects to address for future work.

\section{Geometry of the parameter space\label{sec:Geometry}}

Before studying the Dicke and LMG models, we briefly review the main features of the QMT and its classical analog. The QMT is a second-rank covariant symmetric tensor which measures the separation in the parameter space between two quantum states with infinitesimally different parameters~\cite{Provost}. Consider a quantum system in the time interval $t\in(-\infty,0)$ which is described through the path integral formulation by a Hamiltonian $H(q(t),p(t);x)$, where $q=\{q^{a}\}$ and $p=\{p_{a}\}$, $a=1,...,n$ are the coordinates and momenta, and $x=\{x^{i}\}$ with $i=1,...,{\cal N}$ is a set of ${\cal N}$ adiabatic parameters. Let us now suppose that at $t=0$, a perturbation is turned on such that during the time interval $t\in(0,\infty)$ the system is described by a perturbed Hamiltonian $H^{\prime}\!=\!H+\mathcal{O}_{i}\delta x^{i}$,  where the deformations $\mathcal{O}_{i}(t)$ are given by
\begin{equation}
\mathcal{O}_{i}(t):=\left(\frac{\partial H}{\partial x^{i}}\right)_{q,p}.\label{defo}
\end{equation}
In order to compare the ground states $|0\rangle$ and $|0^{\prime}\rangle$ that belong to the systems described by $H$ and $H^{\prime}$, respectively, we introduce the fidelity. It is defined as ${\cal F}(x,x+\delta x)=|\langle0^{\prime}|0\rangle|$ and its expansion to second order in $\delta x^{i}$ yields ${\cal F}(x,x+\delta x)=1-\frac{1}{2}g_{ij}^{(0)}(x)\delta x^{i}\delta x^{j}$, where $g_{ij}^{(0)}(x)$ is the QMT for the ground state and reads as~\cite{Alvarez2017}
\begin{align}
g_{ij}^{(0)}(x)=&-\frac{1}{\hbar^{2}}\intop_{-\infty}^{0}dt_{1}\intop_{0}^{\infty}dt_{2}\,\bigg(\frac{1}{2}\langle\{\hat{{\cal O}}_{i}(t_{1}),\hat{{\cal O}}_{j}(t_{2})\}\rangle_{0} \nonumber \\
&-\langle\hat{{\cal O}}_{i}(t_{1})\rangle_{0}\langle\hat{{\cal O}}_{j}(t_{2})\rangle_{0}\bigg).\label{qmetric}
\end{align}
In this expression, $\hat{{\cal O}}_{i}(t)$ are the Heisenberg operators corresponding to the deformations~(\ref{defo}), which can be written now as
\begin{equation}
\hat{{\cal O}}_{i}(t)=\left(\frac{\partial\hat{H}}{\partial x^{i}}\right)_{\hat{q},\hat{p}},
\end{equation}
and are functions of $\hat{q}(t)$, $\hat{p}(t)$ and the parameters $x$. Also, the symbol $\{\cdot,\cdot\}$ stands for the anticommutator of two operators, and the expectation values, denoted as $\langle\cdot\rangle_{0}$, are taken in the ground state of the system with Hamiltonian $\hat{H}$. It is worth mentioning that this approach to the QMT can be naturally adapted to problems where the ground state is not known and perturbation theory is required~\cite{AlvarezGenerating}.

The QMT possesses a classical analog, called the classical metric tensor, which measures the distance in parameter space between the phase space points $(q(x),p(x))$ and $(q(x+\delta x),p(x+\delta x))$. It can be shown to arise from the semiclassical approximation of Eq.~(\ref{qmetric}) for integrable systems~\cite{GonzalezAnnalen} and is given by
\begin{align}
g_{ij}(x)=&-\intop_{-\infty}^{0}dt_{1}\intop_{0}^{\infty}dt_{2}\,\big(\langle{\cal O}_{i}(t_{1}){\cal O}_{j}(t_{2})\rangle_{\mathrm{cl}} \nonumber \\
&-\langle{\cal O}_{i}(t_{1})\rangle_{\mathrm{cl}}\langle{\cal O}_{j}(t_{2})\rangle_{\mathrm{cl}}\big),\label{classDef}
\end{align}
where the ${\cal O}_{i}(t)$ are the classical deformation functions given by
\begin{equation}
{\cal O}_{i}(t)=\left(\frac{\partial H}{\partial x^{i}}\right)_{q,p}.\label{fucntionO}
\end{equation}
These functions can be written in terms of the initial conditions $(q_{0},p_{0})$ and time and,  subsequently, in terms of the initial action-angle variables $(\phi_{0},I)$ and time; therefore, the deformations (\ref{fucntionO}) with their full dependence are ${\cal O}_{i}(t)={\cal O}_{i}(q(\phi_{0},I,t;x),p(\phi_{0},I,t;x);x)$. The notation $\langle f\rangle_{\mathrm{cl}}$ stands for the classical torus average of the function $f(\phi_{0},I,t;x)$ over the $n$ initial angle variables,
\begin{equation}
\langle f\rangle_{\mathrm{cl}}=\frac{1}{(2\pi)^{n}}\intop_{0}^{2\pi}d^{n}\phi_{0}\,f(\phi_{0},I,t;x).\label{classAv}
\end{equation}
Notice that this classical average replaces the quantum expectation value that appears in Eq.~(\ref{qmetric}).

Now that we have endowed our parameter space with a metric structure, we can construct a quantity which in two dimensions contains all the manifold's local information in an invariant way: the scalar curvature, also known as the Ricci scalar. In two dimensions, given the coordinates $(x^{1},x^{2})$, the scalar curvature can be computed as~\cite{Sokolnikoff}
\begin{subequations}\label{defscalar}
\begin{equation}
R=\frac{1}{\sqrt{g}}({\cal A}+{\cal B}),
\tag{\ref{defscalar}}
\end{equation}
where
\begin{align}
{\cal A} & =\frac{\partial}{\partial x^{1}}\left[\frac{1}{\sqrt{g}}\left(\frac{g_{12}}{g_{11}}\frac{\partial g_{11}}{\partial x^{2}}-\frac{\partial g_{22}}{\partial x^{1}}\right)\right], \\
{\cal B} & =\frac{\partial}{\partial x^{2}}\left[\frac{1}{\sqrt{g}}\left(2\frac{\partial g_{12}}{\partial x^{1}}-\frac{\partial g_{11}}{\partial x^{2}}-\frac{g_{12}}{g_{11}}\frac{\partial g_{11}}{\partial x^{1}}\right)\right],
\end{align}
\end{subequations}
and $g$ is the determinant of the metric. Notice that the definition~(\ref{defscalar}) differs by a global sign from that used in Ref.~\cite{Sarkar2012} since we are employing the more common contraction $R:=g^{ij}R^{k}{}_{ikj}$, where $R^{i}{}_{jkl}$  is the Riemann tensor.

Having presented an overview of the parameter space's geometry, we devote the following sections to study the classical metric and its scalar curvature for the Dicke and LMG models and compare them with the results of the quantum analysis.

\section{Dicke model\label{sec:Dicke}}

The Dicke model~\cite{Dicke} describes a collection of $N$ two-level atoms interacting with one mode of a bosonic field inside a cavity. Its quantum and classical dynamics have been explored~\cite{Chen2008,*Chen2009,Bakemeier2013}, and it has been widely studied in the context of quantum and classical chaos~\cite{EmaryPRL,*EmaryPRE,Hirsch2014-1,*Hirsch2014-2}, entanglement and fidelity~\cite{EmaryEntPRL,*EmaryEntPRA,Wang2014,Lewis2019,Hirsch2020}.

The Hamiltonian of the Dicke model is
\begin{equation}
\hat{H}=\omega_{0}\hat{J}_{z}+\omega\hat{a}^{\dagger}\hat{a}+\frac{\lambda}{\sqrt{N}}(\hat{a}^{\dagger}+\hat{a})(\hat{J}_{+}+\hat{J}_{-}),\label{Dickeoriginal}
\end{equation}
where $\omega_{0}$ is the splitting of the two levels, $\omega$ is the frequency of the bosonic mode, $\lambda$ is the coupling of the dipole interaction between the field and the atoms, $\hat{a}$ and $\hat{a}^{\dagger}$ are the creation and annihilation operators of the field, and $\hat{J}_{z},\hat{J}_{\pm}=\hat{J}_{x}\pm i\hat{J}_{y}$ are the collective spin operators. Also, we have chosen $\hbar=1$. We see that the operator $\hat{J}^{2}=\hat{J}_{x}^{2}+\hat{J}_{y}^{2}+\hat{J}_{z}^{2}$ commutes with the Hamiltonian so the total pseudospin is conserved, and we can restrict ourselves, as usual, to the consideration of the maximum pseudospin $j=N/2$. This has the effect of treating the collection of $N$ two-level atoms as a single $(N+1)$-level system with pseudospin $j=N/2$~\cite{EmaryPRL,EmaryPRE}. We are interested in the thermodynamic limit $j\rightarrow\infty$, where the system undergoes a QPT at the critical coupling $\lambda=\lambda_{c}\equiv\frac{\sqrt{\omega\omega_{0}}}{2}$ that separates the normal phase $\lambda<\lambda_{c}$ and the superradiant phase $\lambda>\lambda_{c}$.

\subsection{Analysis in the thermodynamic limit}

\subsubsection{\textbf{Normal phase}}

To describe the system in the thermodynamic limit, we follow the work of Emary and Brandes~\cite{EmaryPRL,EmaryPRE}. We first use the Holstein-Primakoff transformation~\cite{HolsteinPrimakoff}
\begin{equation}
\hat{J}_{+}=\hat{b}^{\dagger}\sqrt{2j-\hat{b}^{\dagger}\hat{b}},\,\,\,\,\,\hat{J}_{-}=\left( \sqrt{2j-\hat{b}^{\dagger}\hat{b}}\right) \hat{b},\,\,\,\,\,\hat{J}_{z}=\hat{b}^{\dagger}\hat{b}-j,
\end{equation}
which is a way to associate the bosonic operators, $\hat{b}$ and $\hat{b}^{\dagger}$, to the angular momentum operators $\hat{J}_{z},\hat{J}_{\pm}$. After performing this transformation, the Dicke Hamiltonian takes the form
\begin{align}
\hat{H}=-j\omega_{0} & +\omega_{0}\hat{b}^{\dagger}\hat{b}+\omega\hat{a}^{\dagger}\hat{a} \nonumber \\
& +\lambda(\hat{a}^{\dagger}+\hat{a})\left(\hat{b}^{\dagger}\sqrt{1-\frac{\hat{b}^{\dagger}\hat{b}}{2j}}+\sqrt{1-\frac{\hat{b}^{\dagger}\hat{b}}{2j}}\,\hat{b}\right).\label{HEmary}
\end{align}
Next, we expand the square roots in Eq.~(\ref{HEmary}) and take the limit $j\rightarrow\infty$, keeping only the zeroth order term in $1/j$. This leads to the effective Hamiltonian
\begin{equation}
\hat{H}_{n}=-j\omega_{0}+\omega_{0}\hat{b}^{\dagger}\hat{b}+\omega\hat{a}^{\dagger}\hat{a}+\lambda(\hat{a}^{\dagger}+\hat{a})(\hat{b}^{\dagger}+\hat{b}),\label{HDickequad}
\end{equation}
which is valid for $\lambda<\lambda_{c}$, i.e., the normal phase. The term proportional to $j$, which is dominant as $j$ increases, is identified as the ground state energy of the system in the normal phase. From Eq.~(\ref{HDickequad}), we readily recognize that this Hamiltonian corresponds to two coupled harmonic oscillators, as can be explicitly seen by applying the operator transformation
\begin{subequations}
\begin{align}
\hat{q}_{1} & =\frac{1}{\sqrt{2\omega}}(\hat{a}^{\dagger}+\hat{a}),\,\,\,\,\,\,\,\,\,\,\hat{p}_{1}=i\sqrt{\frac{\omega}{2}}(\hat{a}^{\dagger}-\hat{a}), \\
\hat{q}_{2} & =\frac{1}{\sqrt{2\omega_{0}}}(\hat{b}^{\dagger}+\hat{b}),\,\,\,\,\,\,\,\,\,\,\hat{p}_{2}=i\sqrt{\frac{\omega_{0}}{2}}(\hat{b}^{\dagger}-\hat{b}),
\end{align}
\end{subequations}
which casts it in the position-momentum representation as
\begin{align}
\hat{H}_{n}=&-j\omega_{0}-\frac{(\omega+\omega_{0})}{2}+\frac{1}{2}\big(\hat{p}_{1}^{2}+\hat{p}_{2}^{2}+\omega^{2}\hat{q}_{1}^{2}+\omega_{0}^{2}\hat{q}_{2}^{2} \nonumber \\
& +4\lambda\sqrt{\omega\omega_{0}}\,\hat{q}_{1}\hat{q}_{2}\big).\label{Hnormal}
\end{align}

We can uncouple the two oscillators by going to the normal coordinates
$(\hat{Q}_{1},\hat{Q}_{2})$ through the transformation
\begin{equation}
\begin{pmatrix}\hat{q}_{1}\\
\hat{q}_{2}
\end{pmatrix}=\begin{pmatrix}\cos\alpha_{n} & \sin\alpha_{n}\\
-\sin\alpha_{n} & \cos\alpha_{n}
\end{pmatrix}\begin{pmatrix}\hat{Q}_{1}\\
\hat{Q}_{2}
\end{pmatrix},\label{transf}
\end{equation}
and similarly for the corresponding conjugate normal momenta $(\hat{P}_{1},\hat{P}_{2})$. The angle $\alpha_{n}$ is such that
\begin{equation}
\tan2\alpha_{n}=\frac{4\lambda\sqrt{\omega\omega_{0}}}{\omega_{0}^{2}-\omega^{2}},\label{anglenormal}
\end{equation}
with $\alpha_{n}\in\left(-\frac{\pi}{4},\frac{\pi}{4}\right)$, and we assume that $\omega_{0}\neq\omega$. After performing this transformation, the Hamiltonian acquires the form
\begin{equation}
\hat{H}_{n}=-j\omega_{0}-\frac{(\omega+\omega_{0})}{2}+\frac{1}{2}\left(\hat{P}_{1}^{2}+\hat{P}_{2}^{2}+\varepsilon_{1n}^{2}\hat{Q}_{1}^{2}+\varepsilon_{2n}^{2}\hat{Q}_{2}^{2}\right),
\end{equation}
where the two (squared) normal frequencies are
\begin{subequations}
\label{norfreq}
\begin{align}
\varepsilon_{1n}^{2} & =\frac{1}{2}\left[\omega^{2}+\omega_{0}^{2}-\sqrt{(\omega^{2}-\omega_{0}^{2})^{2}+16\lambda^{2}\omega\omega_{0}}\right], \\
\varepsilon_{2n}^{2} & =\frac{1}{2}\left[\omega^{2}+\omega_{0}^{2}+\sqrt{(\omega^{2}-\omega_{0}^{2})^{2}+16\lambda^{2}\omega\omega_{0}}\right].
\end{align}
\end{subequations}
We clearly see that at the critical coupling $\lambda_{c}=\frac{\sqrt{\omega\omega_{0}}}{2}$, the normal frequency $\varepsilon_{1n}$ vanishes and the system reduces effectively to only one normal mode.

\subsubsection{\textbf{Superradiant phase}}

In the case of the superradiant phase $(\lambda>\lambda_{c})$, one can derive an effective Hamiltonian $\hat{H}_{s}$ by letting the field and the set of atoms acquire macroscopic occupation numbers; one way to achieve this is by displacing the bosonic operators that appear in~(\ref{HEmary}) and demanding that the linear terms in $\hat{a}$ and $\hat{a}^{\dagger}$ vanish. After expanding the square roots and changing to the position-momentum representation, we arrive at the Hamiltonian for the superradiant phase, which reads as~\cite{EmaryPRL,EmaryPRE}
\begin{align}
\hat{H}_{s}= & -j\left(\frac{2\lambda^{2}}{\omega}+\frac{\omega_{0}^{2}\omega}{8\lambda^{2}}\right)-\frac{4\lambda^{2}+\omega^{2}}{2\omega}\nonumber \\
 & +\frac{1}{2}\left(\hat{p}_{1}^{2}+\hat{p}_{2}^{2}+\omega^{2}\hat{q}_{1}^{2}+\frac{16\lambda^{4}}{\omega^{2}}\hat{q}_{2}^{2}+2\omega\omega_{0}\,\hat{q}_{1}\hat{q}_{2}\right).\label{Hsuper}
\end{align}
As in the normal phase, we use the transformation~(\ref{transf}), which casts the Hamiltonian in the form
\begin{align}
\hat{H}_{s}= & -j\left(\frac{2\lambda^{2}}{\omega}+\frac{\omega_{0}^{2}\omega}{8\lambda^{2}}\right)-\frac{4\lambda^{2}+\omega^{2}}{2\omega} \nonumber \\
& +\frac{1}{2}\left(\hat{P}_{1}^{2}+\hat{P}_{2}^{2}+\varepsilon_{1s}^{2}\hat{Q}_{1}^{2}+\varepsilon_{2s}^{2}\hat{Q}_{2}^{2}\right),
\end{align}
where now the rotation angle $\alpha_{s}$ is such that
\begin{equation}
\tan2\alpha_{s}=\frac{2\omega^{3}\omega_{0}}{16\lambda^{4}-\omega^{4}},\label{anglesuper}
\end{equation}
and we have assumed that $\lambda\neq\pm\omega/2$. The two resulting (squared) normal frequencies are
\begin{subequations}
\label{freqsuper}
\begin{align}
\varepsilon_{1s}^{2} & =\frac{1}{2}\!\!\left[\frac{16\lambda^{4}+\omega^{4}}{\omega^{2}}-\sqrt{\left(\frac{16\lambda^{4}-\omega^{4}}{\omega^{2}}\right)^{2}\!\!+4\omega^{2}\omega_{0}^{2}}\right], \\
\varepsilon_{2s}^{2} & =\frac{1}{2}\!\!\left[\frac{16\lambda^{4}+\omega^{4}}{\omega^{2}}+\sqrt{\left(\frac{16\lambda^{4}-\omega^{4}}{\omega^{2}}\right)^{2}\!\!+4\omega^{2}\omega_{0}^{2}}\right].
\end{align}
\end{subequations}

Notice, once more, that at the critical coupling $\lambda=\lambda_{c}$, the frequency $\varepsilon_{1s}$ vanishes. Indeed, it can be easily verified that $\hat{H}_{n}(\lambda_{c})=\hat{H}_{s}(\lambda_{c})$. Furthermore, looking at the dominant term of order $j$ in the Hamiltonians~(\ref{Hnormal}) and~(\ref{Hsuper}), we can read off the ground state energy for both phases:
\begin{equation}
\frac{E_{g}}{j}=\begin{cases}
-\omega_{0}, & \lambda<\lambda_{c}\\
-\left(\frac{2\lambda^{2}}{\omega}+\frac{\omega_{0}^{2}\omega}{8\lambda^{2}}\right), & \lambda>\lambda_{c}.
\end{cases}
\end{equation}
This normalized ground state energy exhibits a discontinuity in its second derivative at $\lambda=\lambda_{c}$, which is precisely the hallmark of the QPT in this model. Interestingly, the main features of the QPT were reproduced using only a quadratic approximation coming from the truncated Holstein-Primakoff transformation. This is one of the main virtues of this approach, which was precisely exploited by Emary and Brandes in their remarkable papers~\cite{EmaryPRL,EmaryPRE}. We are also taking advantage of this method as a first step toward understanding the underlying
geometry of the parameter space. The interested reader can consult Ref.~\cite{Hirsch2011}, where some shortcomings of the truncated Holstein-Primakoff approximation are addressed.

\subsection{Classical and quantum metric tensors for the Dicke model\label{subsec:MetricDicke}}

Our aim is now to compute the classical and quantum metrics for the normal and superradiant phases and compare them to see how well the classical metric captures the essential information of the quantum system. After that, we will analyze the scalar curvatures of both metrics. We fix $\omega_{0}={\rm const}$ and take as our adiabatic parameters the frequency $\omega$ and the strength of the dipole coupling $\lambda$, which results in a two-dimensional parameter manifold with coordinates $x=\{x^{i}\}=(\omega,\lambda),\,i=1,2$.

\subsubsection{\textbf{Metrics of the normal phase}}

We begin our computation of the classical metric tensor~(\ref{classDef}) for the normal phase, whose Hamiltonian is the classical counterpart of Eq.~(\ref{Hnormal}):
\begin{align}\label{Hamno}
H_{n,cl}=&-j\omega_{0}-\frac{(\omega+\omega_{0})}{2}+\frac{1}{2}\big(p_{1}^{2}+p_{2}^{2}+\omega^{2}q_{1}^{2}+\omega_{0}^{2}q_{2}^{2} \nonumber \\
& +\!4\lambda\sqrt{\omega\omega_{0}}\,q_{1}q_{2}\!\big).
\end{align}
The deformation functions associated to the parameters are
\begin{subequations}
\begin{align}
{\cal O}_{1n} & =\frac{\partial H_{n,cl}}{\partial\omega}=\omega q_{1}^{2}+\lambda\sqrt{\frac{\omega_{0}}{\omega}}\,q_{1}q_{2}, \\
{\cal O}_{2n} & =\frac{\partial H_{n,cl}}{\partial\lambda}=2\sqrt{\omega\omega_{0}}\,q_{1}q_{2},
\end{align}
\end{subequations}
where we have ignored the terms that do not depend on $(q_{a},p_{a})$ since they would not contribute to the metric integrands $\Lambda_{ij}(t_{1},t_{2}):=\langle{\cal O}_{in}(t_{1}){\cal O}_{jn}(t_{2})\rangle_{\mathrm{cl}}-\langle{\cal O}_{in}(t_{1})\rangle_{\mathrm{cl}}\langle{\cal O}_{jn}(t_{2})\rangle_{\mathrm{cl}}$.
Actually, we can deal with both deformation functions simultaneously and write them as
\begin{align}
{\cal O}_{in}(t)=&\varepsilon_{1n}Q_{1}^{2}(t)\partial_{i}\varepsilon_{1n}+\varepsilon_{2n}Q_{2}^{2}(t)\partial_{i}\varepsilon_{2n} \nonumber \\
& +(\varepsilon_{2n}^{2}-\varepsilon_{1n}^{2})Q_{1}(t)Q_{2}(t)\partial_{i}\alpha_{n},\label{deffunct}
\end{align}
where $\partial_{i}:=\partial/\partial x^{i}$, the $Q_{a}(t),\,\,(a=1,2)$ are the normal coordinates that uncouple the two harmonic oscillators through the transformation~(\ref{transf}), and $\varepsilon_{an}$ are the normal frequencies~(\ref{norfreq}). The next step is to write the $Q_{a}(t)$ as functions of the initial conditions $(Q_{a0},P_{a0})$ and time as
\begin{equation}
Q_{a}(t)=Q_{a0}\cos\varepsilon_{an}t+\frac{P_{a0}}{\varepsilon_{an}}\sin\varepsilon_{an}t,
\end{equation}
and then, the initial conditions in terms of initial action-angle variables $(\phi_{a0},I_{a})$ as
\begin{equation}
Q_{a0}=\sqrt{\frac{2I_{a}}{\varepsilon_{an}}}\sin\phi_{a0},\,\,\,\,\,\,\,\,\,\,P_{a0}=\sqrt{2I_{a}\varepsilon_{an}}\cos\phi_{a0}.
\end{equation}

Now, we use the classical torus average~(\ref{classAv}) to form the integrands $\Lambda_{ij}(t_{1},t_{2})$ which turn out to be
\begin{align}
\Lambda_{ij}(t_{1},t_{2})=& \frac{1}{2}\partial_{i}\varepsilon_{1n}\partial_{j}\varepsilon_{1n}I_{1}\cos\left(2\varepsilon_{1n}T\right) \nonumber \\
& +\frac{1}{2}\partial_{i}\varepsilon_{2n}\partial_{j}\varepsilon_{2n}I_{2}\cos\left(2\varepsilon_{2n}T\right) \nonumber \\
& +\frac{\partial_{i}\alpha_{n}\partial_{j}\alpha_{n}}{\varepsilon_{1n}\varepsilon_{2n}}\!\left(\!\varepsilon_{1n}^{2}\!-\varepsilon_{2n}^{2}\!\right)^{2}\!\cos\left(2\varepsilon_{1n}T\right)\cos\left(2\varepsilon_{2n}T\right).\label{integrandsDicke}
\end{align}
where $T=t_1-t_2$. Then, we convert the trigonometric functions to complex exponentials, substitute~(\ref{integrandsDicke}) into~(\ref{classDef}), and use the standard regularization
\begin{align}
\intop_{-\infty}^{0}dt_{1}\intop_{0}^{\infty}dt_{2}\,e^{\pm i\Omega T} & :=\lim_{\delta\rightarrow0^{+}}\intop_{-\infty}^{0}dt_{1}\intop_{0}^{\infty}dt_{2}\,e^{\pm i(\Omega\mp i\delta)T} \nonumber \\
& =-\frac{1}{\Omega^{2}}\label{reg}
\end{align}
to finally obtain the classical metric for the normal phase, whose components are
\begin{align}
g_{ij}=&\frac{\partial_{i}\varepsilon_{1n}\partial_{j}\varepsilon_{1n}}{8\varepsilon_{1n}^{2}}I_{1}^{2}+\frac{\partial_{i}\varepsilon_{2n}\partial_{j}\varepsilon_{2n}}{8\varepsilon_{2n}^{2}}I_{2}^{2} \nonumber \\
& +\partial_{i}\alpha_{n}\partial_{j}\alpha_{n}\left(\frac{\varepsilon_{1n}}{\varepsilon_{2n}}+\frac{\varepsilon_{2n}}{\varepsilon_{1n}}\right)I_{1}I_{2}.\label{classmetric}
\end{align}
It  is  clear  from this expression that the appearance of normal frequency $\varepsilon_{1n}$ in the denominator causes a divergence in the metric components at $\lambda_{c}=\frac{\sqrt{\omega\omega_{0}}}{2}$ since $\varepsilon_{1n}$ vanishes at the critical coupling [see Eq.~(\ref{norfreq})]; this property of the classical metric signals the QPT in the Dicke model.

We now compute the QMT~(\ref{qmetric}) for the normal phase. From the Hamiltonian (\ref{Hnormal}), we obtain the corresponding deformation operators, which can be written in compact form as
\begin{align}
\hat{{\cal O}}_{in}(t)=&\varepsilon_{1n}\hat{Q}_{1}^{2}(t)\partial_{i}\varepsilon_{1n}+\varepsilon_{2n}\hat{Q}_{2}^{2}(t)\partial_{i}\varepsilon_{2n} \nonumber \\
& +(\varepsilon_{2n}^{2}-\varepsilon_{1n}^{2})\hat{Q}_{1}(t)\hat{Q}_{2}(t)\partial_{i}\alpha_{n},
\end{align}
where $\hat{Q}_{a}(t),\,\,(a=1,2)$ are the operators corresponding to the normal modes of the diagonal Hamiltonian, which can be written in terms of annihilation and creation operators as
\begin{equation}
\hat{Q}_{a}(t)=\frac{1}{\sqrt{2\varepsilon_{an}}}\left(\hat{b}_{a0}^{\dagger}e^{i\varepsilon_{an}t}+\hat{b}_{a0}e^{-i\varepsilon_{an}t}\right).\label{QHeis}
\end{equation}
With these operators at hand, we can compute the combination $\frac{1}{2}\langle\{\hat{{\cal O}}_{in}(t_{1}),\hat{{\cal O}}_{jn}(t_{2})\}\rangle_{0}-\langle\hat{{\cal O}}_{in}(t_{1})\rangle_{0}\langle\hat{{\cal O}}_{jn}(t_{2})\rangle_{0}$ and then use the regularization~(\ref{reg}) to arrive at the components of the QMT,  which turn out to be
\begin{align}
g_{ij}^{(0)}=&\frac{\partial_{i}\varepsilon_{1n}\partial_{j}\varepsilon_{1n}}{8\varepsilon_{1n}^{2}}+\frac{\partial_{i}\varepsilon_{2n}\partial_{j}\varepsilon_{2n}}{8\varepsilon_{2n}^{2}} \nonumber \\
& +\partial_{i}\alpha_{n}\partial_{j}\alpha_{n}\left[\frac{1}{4}\left(\frac{\varepsilon_{1n}}{\varepsilon_{2n}}+\frac{\varepsilon_{2n}}{\varepsilon_{1n}}\right)-\frac{1}{2}\right].\label{quantummetric}
\end{align}
Notice that, as in the case of the classical metric, the frequency $\varepsilon_{1n}$ appears in the denominator in Eq.~(\ref{quantummetric}), causing a divergence when $\lambda=\lambda_{c}$ and signaling the QPT in the Dicke model. Moreover, with both metrics [Eqs.~(\ref{classmetric}) and (\ref{quantummetric})] at our disposal, we find the relation
\begin{equation}
g_{ij}^{(0)}=g_{ij}-\frac{1}{2}\partial_{i}\alpha_{n}\partial_{j}\alpha_{n},\label{relQuantClass}
\end{equation}
where we have made the identifications $I_{1}=I_{2}=1/2$ and $I_{1}^{2}=I_{2}^{2}=1$, which will be assumed in the rest of this paper. From (\ref{relQuantClass}) it is clear that the QMT~(\ref{quantummetric}) has an extra parameter-dependent term that does not appear in its classical analog~(\ref{classmetric}); these type of terms have been related to an anomaly arising from the ordering of the operators in the quantum case (see Ref.~\cite{GonzalezAnnalen} for details).

\subsubsection{\textbf{Metrics of the superradiant phase}}

\begin{figure}[ht]
\includegraphics[width=\columnwidth]{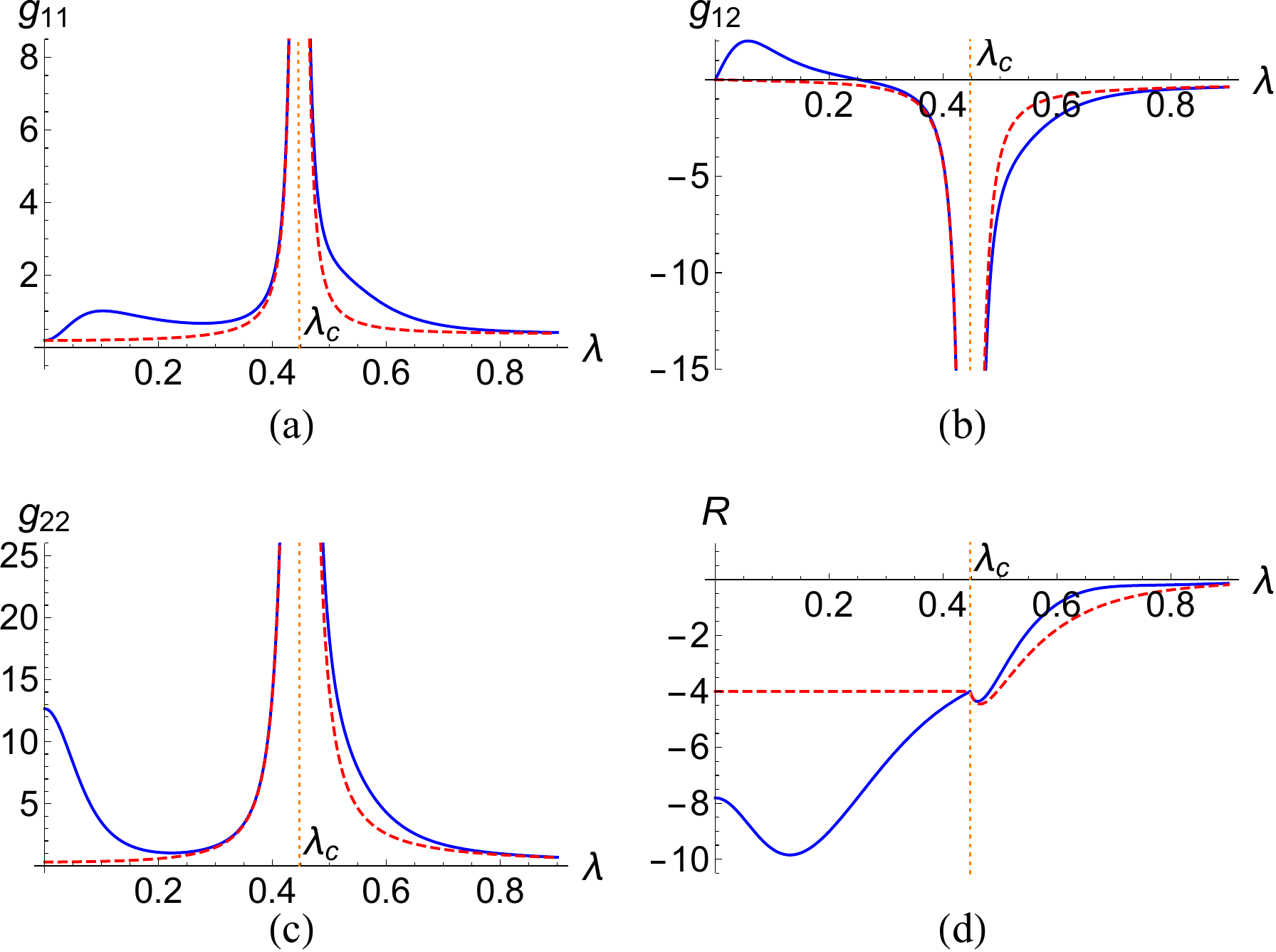}
\caption{Metric components and scalar curvature of the classical metric (solid blue) and the quantum metric (dashed red) as a function of $\lambda$ when $\omega_0=1$ and $\omega=0.8$. All the components show a divergence at the phase transition (dotted orange) with critical coupling $\lambda_c=0.447$, whereas the scalar curvature does not.}
\label{FigNo_res}
\end{figure}

The treatment of the superradiant phase is analogous to that of the normal phase in both the classical and quantum settings; the difference lies in the explicit expressions of the rotation angle~(\ref{anglesuper}) and the normal frequencies~(\ref{freqsuper}) in terms of the parameters. The classical counterpart of the Hamiltonian~(\ref{Hsuper}) is
\begin{align}
H_{s,cl}= & -j\left(\frac{2\lambda^{2}}{\omega}+\frac{\omega_{0}^{2}\omega}{8\lambda^{2}}\right)-\frac{4\lambda^{2}+\omega^{2}}{2\omega} \nonumber \\
 & +\frac{1}{2}\left(p_{1}^{2}+p_{2}^{2}+\omega^{2}q_{1}^{2}+\frac{16\lambda^{4}}{\omega^{2}}q_{2}^{2}+2\omega\omega_{0}\,q_{1}q_{2}\right),
\end{align}
and its deformation functions are
\begin{subequations}
\begin{align}
{\cal O}_{1s} & =\frac{\partial H_{s,cl}}{\partial\omega}=\omega q_{1}^{2}-\frac{16\lambda^{4}}{\omega^{3}}q_{2}^{2}+\omega_{0}\,q_{1}q_{2}, \\
{\cal O}_{2s} & =\frac{\partial H_{s,cl}}{\partial\lambda}=\frac{32\lambda^{3}}{\omega^{2}}q_{2}^{2}.
\end{align}
\end{subequations}
By following the same steps as in the normal phase, we arrive at the classical and quantum metrics,  which turn out to be
\begin{align}
	g_{ij}=&\frac{\partial_{i}\varepsilon_{1s}\partial_{j}\varepsilon_{1s}}{8\varepsilon_{1s}^{2}}I_{1}^{2}+\frac{\partial_{i}\varepsilon_{2s}\partial_{j}\varepsilon_{2s}}{8\varepsilon_{2s}^{2}}I_{2}^{2} \nonumber \\
	& +\partial_{i}\alpha_{s}\partial_{j}\alpha_{s}\left(\frac{\varepsilon_{1s}}{\varepsilon_{2s}}+\frac{\varepsilon_{2s}}{\varepsilon_{1s}}\right)I_{1}I_{2}\label{classmetricS}
\end{align}	
and
\begin{align}
	g_{ij}^{(0)}=&\frac{\partial_{i}\varepsilon_{1s}\partial_{j}\varepsilon_{1s}}{8\varepsilon_{1s}^{2}}+\frac{\partial_{i}\varepsilon_{2s}\partial_{j}\varepsilon_{2s}}{8\varepsilon_{2s}^{2}} \nonumber \\
	& +\partial_{i}\alpha_{s}\partial_{j}\alpha_{s}\left[\frac{1}{4}\left(\frac{\varepsilon_{1s}}{\varepsilon_{2s}}+\frac{\varepsilon_{2s}}{\varepsilon_{1s}}\right)-\frac{1}{2}\right],\label{quantummetricS}
\end{align}	
respectively. Notice that these metrics have the same form as those of Eqs.~(\ref{classmetric}) and~(\ref{quantummetric}) and hence satisfy the relation~(\ref{relQuantClass}), but with the normal frequencies $\varepsilon_{1s}$ and $\varepsilon_{2s}$ and the rotation angle $\alpha_{s}$, showing a quantum anomaly effect. It is worth mentioning that despite their same appearance, the metrics are entirely different when written explicitly in terms of the Hamiltonian parameters $x=(\omega,\lambda)$. This fact accounts for their different form when plotted, as can be seen in Fig.~\ref{FigNo_res}. Moreover, due to the presence of $\varepsilon_{1s}$ in the denominator of (\ref{classmetricS}) and (\ref{quantummetricS}), both metrics exhibit a divergence at $\lambda=\lambda_{c}$, which reveals the existence of the QPT. Remarkably, this shows once again that the classical metric is sensitive to the presence of the QPT.

To gain more insight into this, in Figs.~\hyperref[FigNo_res]{1(a)--1(c)} we show the components of the classical and quantum metrics for both phases and fixed values of $\omega$ and $\omega_{0}$. Clearly, we see that both metrics diverge at $\lambda_{c}$, which signals the QPT. Moreover, these metrics have a very close behavior in the neighborhood of $\lambda_{c}$. Nevertheless, the component $g_{22}$ of the classical metric at $\lambda=0$ shows a different behavior than its quantum counterpart. This can be attributed to the fact that the anomalous extra term that appears in~(\ref{quantummetricS}) produces this notorious difference, making the classical metric more sensitive to the vanishing of the coupling term in the Hamiltonian (\ref{Hamno}). The scalar curvatures, computed using (\ref{defscalar}), for both the classical and quantum metrics are shown in Fig.~\hyperref[FigNo_res]{1(d)}. We observe some features from the plots. First, the agreement between them is right in the superradiant phase $(\lambda>\lambda_{c})$. Second, an important difference between them appears in the normal phase ($\lambda<\lambda_{c}$): while in the quantum case the scalar curvature takes a constant value very close to $-4$,  in the classical case the scalar curvature possesses a minimum around $\lambda=0.16$.  Such a difference can be related to the behavior of the component $g_{22}$ at that phase. And third, in the limit $\lambda\rightarrow\lambda_{c}$ the scalar curvatures approach each other and tend to $-4$. Notice, however, that for $\lambda=\lambda_{c}$ they are not defined since the classical and quantum Hamiltonians used to compute the metrics are not valid at that point. The behavior of the scalar curvatures when $\lambda\rightarrow\lambda_{c}$ implies that the singularity of the classical and quantum metrics at the QPT is merely an artifact of the parameter space's coordinates, i.e., it is a removable singularity. This is why in the next model that we analyze, we carry out a numerical study for finite $j$ to elucidate the nature of the singularity. It is worth mentioning that the scalar curvature of the quantum metric resembles the one found in Ref.~\cite{Sarkar2012}, the only difference being the method used to compute it and the overall sign due to the convention that we explained in Sec.~\ref{sec:Geometry}.

\subsubsection{\textbf{Metrics under resonance}}

A special important case of the Dicke model is that of resonance, i.e., when $\omega=\omega_{0}$. We are unaware of previous geometric analyses in this case. In order to treat it, we take the limits $\alpha_{n}\rightarrow\pi/4$ and $\omega_{0}\rightarrow\omega$ in the classical and quantum metrics corresponding to the normal phase [Eqs.~(\ref{classmetric}) and (\ref{quantummetric})], whereas in the superradiant phase we only set $\omega_{0}\rightarrow\omega$ in the associated metrics [Eqs.~(\ref{classmetricS}) and~(\ref{quantummetricS})]. The resulting metric components in terms of the parameters are greatly simplified, and we find that in the normal phase, the components $g_{12}$ and $g_{22}$ of the classical and quantum metrics perfectly match when the identifications of the action variables are used. These components are
\begin{subequations}
\begin{align}
g_{12} = g_{12}^{(0)}  & =\frac{\lambda(4\lambda^{2}-3\omega^{2})}{8\omega(\omega^{2}-4\lambda^{2})^{2}}, \\
g_{22} = g_{22}^{(0)} & =\frac{4\lambda^{2}+\omega^{2}}{4(\omega^{2}-4\lambda^{2})^{2}}.
\end{align}
\end{subequations}
On the other hand, the $g_{11}$ and $g_{11}^{(0)}$ components do not match, as we can see below:

\begin{widetext}
\begin{subequations}
\begin{align}
g_{11}& =\frac{16\lambda^{4}\omega^{3}-8\lambda^{2}\omega^{5}+\omega^{7}+\lambda^{2}\sqrt{\omega^{2}-4\lambda^{2}}(8\lambda^{4}-6\lambda^{2}\omega^{2}+2\omega^{4})}{32\lambda^{2}\omega^{2}\left(\omega^{2}-4\lambda^{2}\right)^{5/2}}, \\
g_{11}^{(0)}& =\frac{-16\lambda^{6}+48\lambda^{4}\omega^{2}-23\lambda^{2}\omega^{4}+3\omega^{6}-\omega\sqrt{\omega^{2}-4\lambda^{2}}(4\lambda^{4}-3\lambda^{2}\omega^{2}+\omega^{4})}{16\omega^{2}(\omega^{2}-4\lambda^{2})^{5/2}\left(\omega+\sqrt{\omega^{2}-4\lambda^{2}}\right)}.
\end{align}
\end{subequations}
\end{widetext}

\begin{figure}[ht]
\includegraphics[width=\columnwidth]{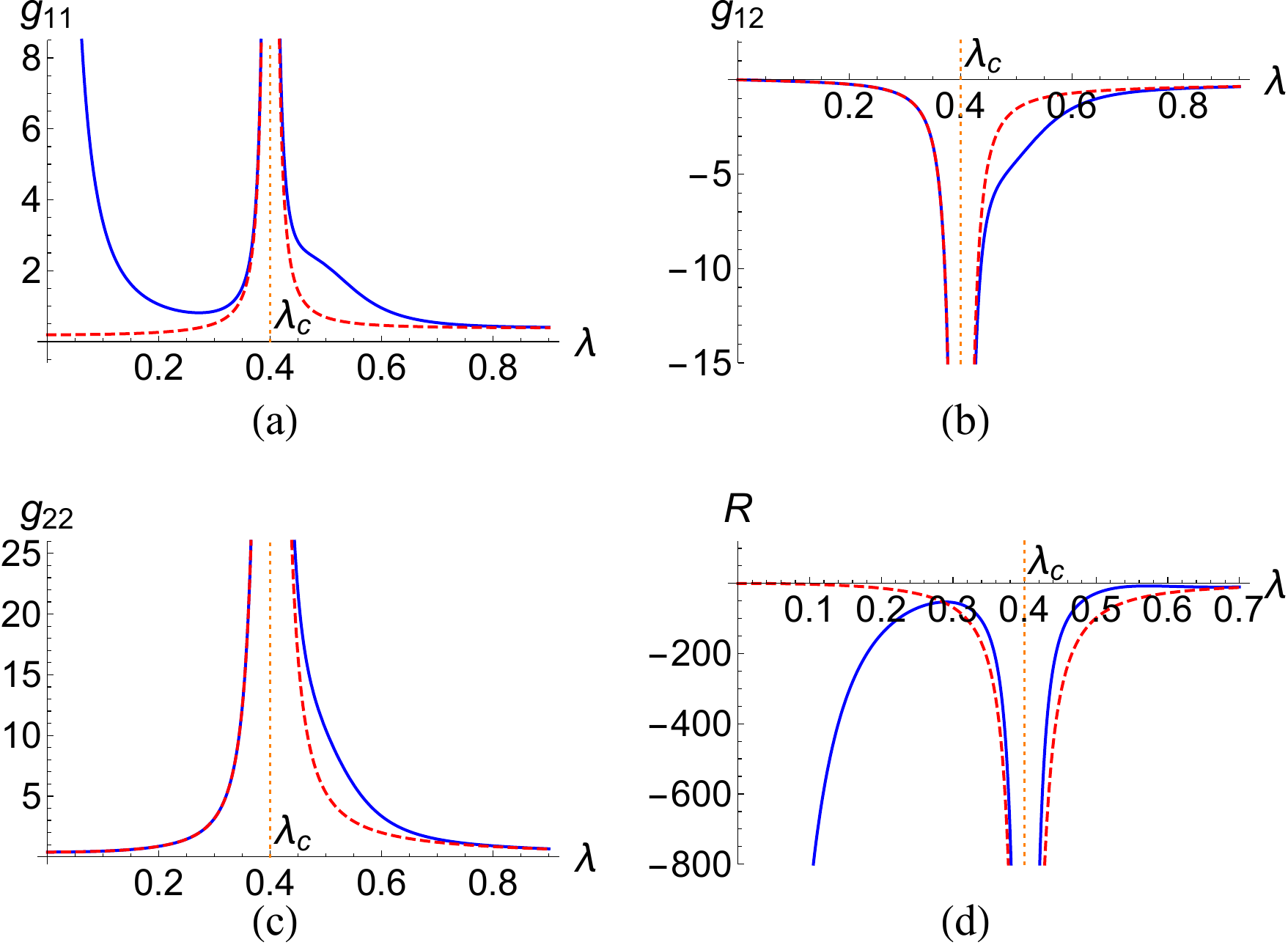}
\caption{Metric components and scalar curvature of the classical metric (solid blue) and the quantum metric (dashed red) for the resonant case as a function of $\lambda$ when $\omega=0.8$. All of them show a divergence at the phase transition (dotted orange) with critical coupling $\lambda_c=0.4$.}
\label{FigRes}
\end{figure}

In the superradiant phase, the classical and quantum metric components are more complicated and do not match; however, it can be seen that all of them diverge at the critical coupling $\lambda_{c}=\omega/2$. We show in Figs.~\hyperref[FigRes]{2(a)--2(c)} the components of the metrics under the resonance condition $\omega=\omega_{0}$, for both phases. We can see that the component $g_{11}$ has a divergence at $\lambda=0$. Moreover, we observe that in the normal phase, the components $g_{12}$ and $g_{22}$ of the classical metric are exactly the same as those of the quantum metric, just as we mentioned earlier. It is also worth noting that both metrics show the same behavior in the limiting cases $\lambda\rightarrow\lambda_{c}$ and $\lambda\rightarrow\infty$.

Figure~\hyperref[FigRes]{2(d)} shows the scalar curvatures associated with the classical and quantum metrics. We notice that the scalar curvature in the classical case presents a divergence at $\lambda=0$, which is inherited from the $g_{11}$ metric component. Once again, we see that the anomaly's role is to get rid of that singularity in the quantum result. In this regard, it might be interesting to study if the behavior of the Dicke model in the resonance case can be considered as a quantum simulator~\cite{Gooding2020Unruh} of some kind of cosmology. Furthermore, in contrast to the nonresonant analysis, both scalar curvatures diverge at the QPT in the same way. There is an alternative approach to the metrics under the resonance condition. One could set $\omega_{0}=\omega$ in the Hamiltonian from the very beginning and, from there, derive the corresponding ${\cal O}_{i}$. The resulting deformation operators are different from those we have used, and lead to different expressions for both classical and quantum metrics as functions of $\omega$ and $\lambda$. As a matter of fact, it turns out that both metrics have zero scalar curvature in the whole range of $\omega$ and $\lambda$, which is not particularly illuminating. We are unaware of the physical reason behind this result and consider that it deserves further analysis.

To conclude this section, we would like to stress the fact that both the classical and quantum metrics exhibit a divergent behavior at the QPT for the nonresonant and resonant cases. This is a remarkable result since it shows that the classical metric can be used to get a first insight into the information contained in the QMT. Additionally, according to the analysis of the (classical and  quantum) scalar curvatures in the resonant case, there is a genuine singularity at the QPT which cannot be removed by a change of parameters, while in the nonresonant case there seems to be a spurious (removable) singularity. This effect could be a consequence of the fact that the Holstein-Primakoff approximation fails at the QPT~\cite{Hirsch2013}. Then, in order to clarify this point it would be valuable to carry out a study for finite $j$ of the parameter space associated to the Dicke model. This model has features that make the study for finite $j$ subtle, although some authors have successfully employed techniques that can be implemented numerically and allow the exploration of the system in various regimes~\cite{Bakemeier2013,Hirsch2014-1,Hirsch2014-2}. In the next section, we study a simpler system with a finite Hilbert space that consists of one degree of freedom: the LMG model. In this case, we shall find a perfect agreement between the classical and quantum descriptions of the parameter space in the thermodynamic limit. Furthermore, we will be able to carry out a numerical analysis of the QMT for finite sizes of the system, providing additional information of interest.

\section{Lipkin-Meshkov-Glick model\label{sec:Lipkin}}

The LMG model consists of $N$ mutually interacting spin-half particles affected by a transverse magnetic field. It was first introduced in the context of nuclear physics~\cite{Lipkin1,Lipkin2,Lipkin3}, and it has been deeply studied through various analytic and numerical techniques~\cite{Vidal2004,*Vidal2005,Leyvraz2005,VidalPRL2007,*VidalPRE2008,Hirsch2013}. Furthermore, it has been used as a model for Floquet time crystals~\cite{Russomanno2017}, in the study of out-of-time order correlators~\cite{Hirsch2020}, and to illustrate the orthogonality catastrophe and its relation to quantum speed limit~\cite{Fogarty2020}. 

The Hamiltonian considered is given by~\cite{GuPRE}
\begin{equation}
\hat{H}=-h\sum_{i}\sigma_{z}^{i}-\frac{1}{N}\sum_{i<j}(\sigma_{x}^{i}\sigma_{x}^{j}+\gamma\sigma_{y}^{i}\sigma_{y}^{j}),
\end{equation}
where $\sigma_{x}^{i}$, $\sigma_{y}^{i}$, and $\sigma_{z}^{i}$ are the Pauli spin matrices for the $i$th spin, $h,\gamma$ are real parameters, and we have set $\hbar=1$. From here, it is customary to define the pseudospin (collective spin) operators as $\hat{J}_{\alpha}=\sum_{i}\sigma_{\alpha}^{i}/2$ and cast the Hamiltonian into the form
\begin{equation}
\hat{H}=-2h\hat{J}_{z}-\frac{1}{j}\left(\hat{J}_{x}^{2}+\gamma\hat{J}_{y}^{2}\right),\label{HLipkin}
\end{equation}
where, as usual, we restrict ourselves to the maximum pseudospin representation with $j=N/2$. We also consider $h\geq0$ and $-1<\gamma<1$, and analyze the system in the thermodynamic limit $j\rightarrow\infty$, where the QPT occurs.

\subsection{Analysis in the thermodynamic limit}

The description of the LMG model when $j\rightarrow\infty$ begins by taking the expectation value of the Hamiltonian~(\ref{HLipkin}) in spin coherent states $|z\rangle$ given by~\cite{Hirsch2006,Hirsch2020}
\begin{equation}
|z\rangle=\frac{e^{z\hat{J}_{+}}}{(1+|z|^{2})^{j}}|j,-j\rangle,
\end{equation}
where $|j,-j\rangle$ is the state with the lowest pseudospin projection and $z$ is a complex number parametrized in terms of the two angles of the Bloch sphere as $z=e^{i\phi}\tan\frac{\theta}{2}$. The function thus obtained is
\begin{equation}
H_{cl}(\theta,\phi):=\lim_{j\rightarrow\infty}\langle z|\hat{H}|z\rangle=-2hJ_{z}-\frac{1}{j}(J_{x}^{2}+\gamma J_{y}^{2}),\label{classLipkin}
\end{equation}
and it defines the classical energy surface where the pseudospin vector $\vec{J}=j(\sin\theta\cos\phi,\sin\theta\sin\phi,\cos\theta)$ dynamics will take place. Explicitly in terms of the angles $(\theta,\phi)$, the function 
\begin{equation}
H_{cl}=-j[2h\cos\theta+\sin^{2}\theta(\cos^{2}\phi+\gamma\sin^{2}\phi)]
\end{equation}
possesses two extrema, each of which defines a phase of the system. These phases are as follows:

(i) Symmetric phase: $\theta_{0}=0$. It corresponds to a classical pseudospin vector aligned with the $z$ axis. The ground state energy is $E_{g}:=H_{cl}(0,\phi_{0})=-2hj$.

(ii) Broken phase: $\theta_{0}=\cos^{-1}h$ with $\phi_{0}=0$ or $\phi_{0}=\pi$. It corresponds to two possible configurations of the pseudospin vector, signaling two ground states with energy $E_{g}:=H_{cl}(\cos^{-1}h,0)=H_{cl}(\cos^{-1}h,\pi)=-(1+h^{2})j$. In this case, the classical pseudospin is not aligned with the $z$ axis.

The ground state energy $E_{g}=E_{g}(h)$ is thus the piecewise function
\begin{equation}
\frac{E_{g}}{j}=\begin{cases}
-(1+h^{2}), & h<1\\
-2h, & h>1
\end{cases}
\end{equation}
which has a discontinuous second derivative at $h=1$, signaling a second order QPT~\cite{Sachdev}. The region $h>1$ corresponds to the symmetric phase where the ground state is unique, whereas the region $h<1$ is the broken phase which has a degenerate ground state energy. We will not pursue further the treatment with coherent states since the QMT resulting from their use provides no valuable information, as can be easily seen from the dependence of the ground state's coordinates $(\theta_0,\phi_0)$ on the parameters $x=(h,\gamma)$.

\subsubsection{\textbf{Symmetric phase}}

To carry out the analysis of the symmetric phase in the thermodynamic limit, we will use again the Holstein-Primakoff transformation~\cite{HolsteinPrimakoff}: 
\begin{equation}
\hat{J}_{-}=\hat{a}^{\dagger}\sqrt{2j-\hat{a}^{\dagger}\hat{a}},\,\,\,\,\,\hat{J}_{+}=\left( \sqrt{2j-\hat{a}^{\dagger}\hat{a}}\right) \hat{a},\,\,\,\,\,\hat{J}_{z}=j-\hat{a}^{\dagger}\hat{a},
\end{equation}
which is then truncated to zeroth order in $1/j$ under the assumption that $j\rightarrow\infty$. Hence, we have
\begin{equation}\label{QyP}
\hat{J}_{-}\simeq\sqrt{2j}\,\hat{a}^{\dagger},\,\,\,\,\,\,\,\,\,\,\hat{J}_{+}\simeq\sqrt{2j}\,\hat{a},\,\,\,\,\,\,\,\,\,\,\hat{J}_{z}\simeq j-\hat{a}^{\dagger}\hat{a}.
\end{equation}

Taking this into account and using $\hat{J}_{\pm}=\hat{J}_{x}\pm i\hat{J}_{y}$, the resulting quadratic Hamiltonian that corresponds to~(\ref{HLipkin}) is
\begin{equation}
\hat{H}\simeq-\frac{1+\gamma}{2}-2hj-(1+\gamma-2h)\hat{a}^{\dagger}\hat{a}-\frac{1-\gamma}{2}\left(\hat{a}^{\dagger2}+\hat{a}^{2}\right),\label{aadagger}
\end{equation}
which, in terms of $\hat{Q}$ and $\hat{P}$, can be written as
\begin{equation}
\hat{H}\simeq-h-2hj+(h-\gamma)\hat{P}^{2}+(h-1)\hat{Q}^{2}.\label{quantQP}
\end{equation}
From~(\ref{aadagger}), it is clear that the Hamiltonian can be diagonalized through the Bogoliubov transformation from operators $(\hat{a},\hat{a}^{\dagger})$ to $(\hat{b},\hat{b}^{\dagger})$ given by
\begin{equation}
\hat{a}=\cosh\alpha\,\hat{b}+\sinh\alpha\,\hat{b}^{\dagger},\,\,\,\,\,\,\,\hat{a}^{\dagger}=\sinh\alpha\,\hat{b}+\cosh\alpha\,\hat{b}^{\dagger},
\end{equation}
with $\tanh2\alpha=\frac{1-\gamma}{2h-\gamma-1}$. By doing this, the Hamiltonian takes the form
\begin{equation}
\hat{H}\simeq-h-2hj+2\sqrt{(h-1)(h-\gamma)}\left(\hat{b}^{\dagger}\hat{b}+\frac{1}{2}\right).
\end{equation}
It is readily noted here that at the phase transition, $h=1$, the frequency of the resulting harmonic oscillator vanishes.

\subsubsection{\textbf{Broken phase}}

In the case of the broken phase, we need to perform a rotation around the $y$ axis to align the $z$ axis with the pseudospin ground state configuration. Hence, we shall transform the operators $(\hat{J}_{x},\hat{J}_{y},\hat{J}_{z})$ to a new set of operators $(\hat{J}_{x}^{\prime},\hat{J}_{y}^{\prime},\hat{J}_{z}^{\prime})$ as
\begin{equation}
\begin{pmatrix}\hat{J}_{x}\\
\hat{J}_{y}\\
\hat{J}_{z}
\end{pmatrix}=\begin{pmatrix}\cos\theta_{0} & 0 & \sin\theta_{0}\\
0 & 1 & 0\\
-\sin\theta_{0} & 0 & \cos\theta_{0}
\end{pmatrix}\begin{pmatrix}\hat{J}_{x}^{\prime}\\
\hat{J}_{y}^{\prime}\\
\hat{J}_{z}^{\prime}
\end{pmatrix},\label{rot}
\end{equation}
where $\cos\theta_{0}=h$ and $\sin\theta_{0}=\sqrt{1-h^{2}}$, which is the ground state configuration that corresponds to $(\theta_{0},\phi_{0})=(\cos^{-1}h,0)$. Thus, the Hamiltonian of the broken phase turns out to be~\cite{Vidal2004,Vidal2005}
\begin{align}
\hat{H}^{\prime}= & -2h^{2}\hat{J}_{z}^{\prime}+2h\sqrt{1-h^{2}}\hat{J}_{x}^{\prime}-\frac{1}{j}(1-h^{2})\hat{J}_{z}^{\prime2}\nonumber \\
 & -\frac{1}{j}\left[h^{2}\hat{J}_{x}^{\prime2}+h\sqrt{1-h^{2}}(\hat{J}_{x}^{\prime}\hat{J}_{z}^{\prime}+\hat{J}_{z}^{\prime}\hat{J}_{x}^{\prime})+\gamma\hat{J}_{y}^{\prime2}\right].\label{Hbroken}
\end{align}
Next, we apply the truncated Holstein-Primakoff transformation~(\ref{QyP}) to these rotated operators to find the quadratic Hamiltonian for the broken phase. The resulting Hamiltonian is given by 
\begin{equation}
\hat{H}^{\prime}\simeq-(1+h^{2})j+(1-\gamma)\hat{P}^{2}+(1-h^{2})\hat{Q}^{2}\label{quantBroken}
\end{equation}
or, in terms of the creation and annihilation operators $\hat{b}$ and $\hat{b}^{\dagger}$,
\begin{equation}
\hat{H}^{\prime}\simeq-(1+h^{2})j+2\sqrt{(1-h^{2})(1-\gamma)}\left(\hat{b}^{\dagger}\hat{b}+\frac{1}{2}\right).
\end{equation}
We observe once more that at the critical point $h=1$, the frequency of the resulting harmonic oscillator vanishes, which signals the QPT. Now that we have at hand the effective quadratic Hamiltonians for both the symmetric and the broken phases, we proceed to compute the classical and quantum metric tensors.

\subsection{Classical and quantum metric tensors for the LMG model\label{subsec:MetricLipkin}}

\begin{figure}[ht]
\includegraphics[width=\columnwidth]{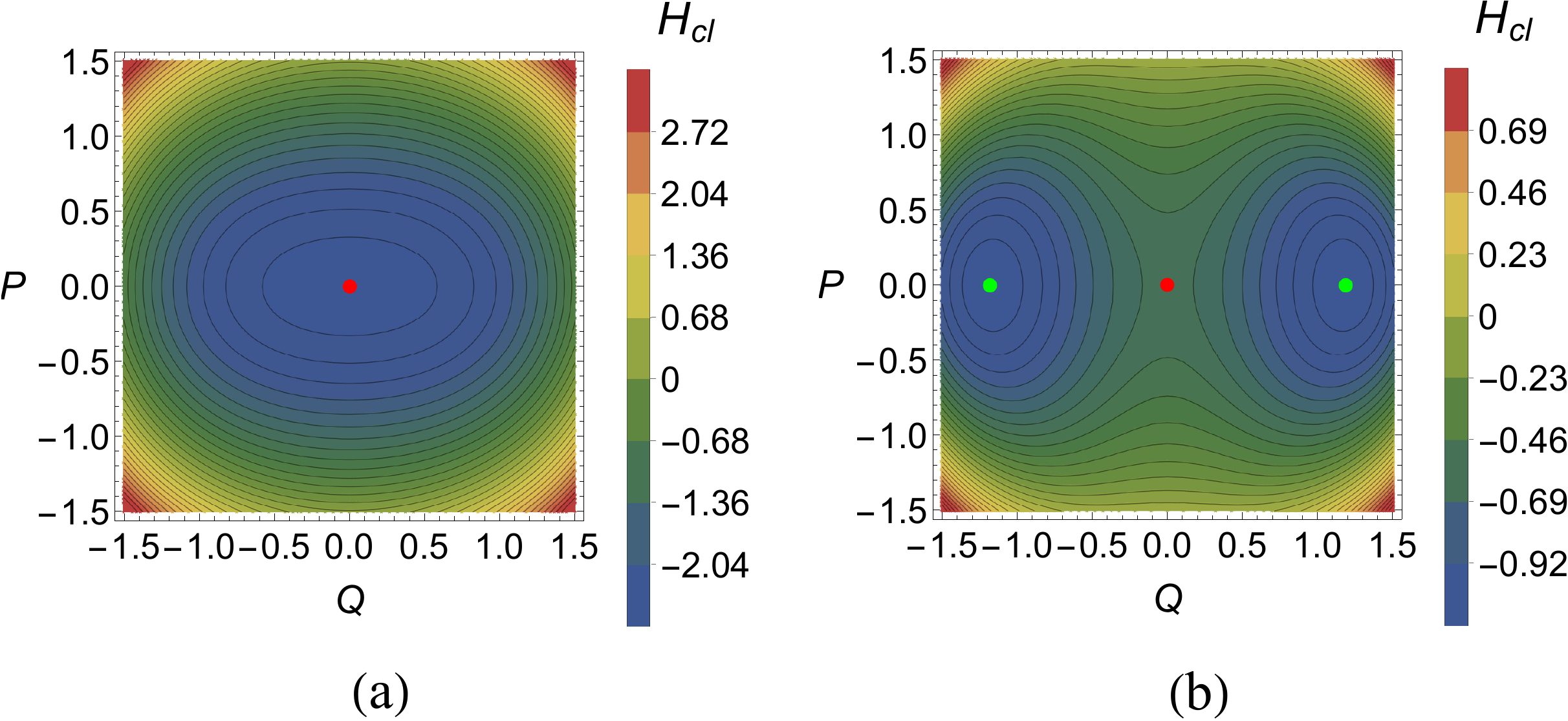}
\caption{Phase space corresponding to the classical Hamiltonian $H_{cl}$ for $\gamma=0.1$ and (a) $h=1.3$ or (b) $h=0.3$. In (a), the red point is the only minimum; this is the symmetric phase. On the other hand, in (b), the red point becomes a local maximum and the green points appear as two degenerate minima; this is the broken phase.}
\label{FigPhase}
\end{figure}

We are now ready to compute the classical and quantum metrics in the thermodynamic limit $j\rightarrow\infty$. In what follows, we take $x=\{x^{i}\}=(h,\gamma),\,i=1,2$, to be the adiabatic parameters. To build the classical metric, we need to derive the deformation functions from the Hamiltonian~(\ref{classLipkin}). They are
\begin{subequations}
\label{defLipkin}
\begin{align}
{\cal O}_{1} & =\frac{\partial H_{cl}}{\partial h}=-2J_{z}, \\
{\cal O}_{2} & =\frac{\partial H_{cl}}{\partial\gamma}=-\frac{J_{y}^{2}}{j}.
\end{align}
\end{subequations}
At this point, we introduce canonical coordinates for the description of the classical system. It is easy to see that the coordinates $(\phi,J_{z})$ are canonical in the sense that they reproduce the angular momentum algebra $\{J_{i},J_{j}\}_{(\phi,J_{z})}=\epsilon_{ijk}J_{k}$, where
\begin{equation}
\{f,g\}_{(\phi,J_{z})}:=\frac{\partial f}{\partial\phi}\frac{\partial g}{\partial J_{z}}-\frac{\partial f}{\partial J_{z}}\frac{\partial g}{\partial\phi}.
\end{equation}
Then, we perform a canonical transformation and move to the $(Q,P)$ representation, where
\begin{equation}
Q=\sqrt{2(j-J_{z})}\cos\phi,\,\,\,\,\,\,\,\,\,\,P=\sqrt{2(j-J_{z})}\sin\phi.\label{transfQPLipkin}
\end{equation}
After this, the resulting classical LMG Hamiltonian is
\begin{equation}
H_{cl}=-2hj\!+\!h(P^{2}\!+\!Q^{2})-(\gamma P^{2}\!+\!Q^{2})\left(1\!-\!\frac{P^{2}\!+\!Q^{2}}{4j}\right).\label{LipkinClassQP}
\end{equation}
In Fig.~\ref{FigPhase}, we show the level curves of the classical Hamiltonian $H_{cl}$ in terms of the $(Q,P)$ coordinates for the two different phases of the model. Once the mean field Hamiltonian $H_{cl}$ is constructed with the coherent states, the analysis is purely classical in terms of fixed points and their stability. This highlights the importance of the classical methods for quantum systems.

\subsubsection{\textbf{Metrics for the symmetric phase}}

We first consider the symmetric phase. The quadratic Hamiltonian associated to~(\ref{LipkinClassQP}) is
\begin{equation}
H_{cl}\simeq-2hj+(h-\gamma)P^{2}+(h-1)Q^{2}.\label{Hsymm}
\end{equation}
Notice that when $Q=P=0$, only the ground state energy of the symmetric phase, $E_{g}=-2hj$, survives. We need to express the deformation functions~(\ref{defLipkin}) in terms of initial action-angle variables and time. To do this, we find the solution to the equations of motion of~(\ref{Hsymm}), which are
\begin{subequations}
\label{solLipkin}
\begin{align}
Q(t)= & Q_{0}\cos\omega t+\frac{P_{0}}{\omega}\sin\omega t, \\
P(t)= & P_{0}\cos\omega t-\omega Q_{0}\sin\omega t,
\end{align}
\end{subequations}
where we have identified the frequency as $\omega=~\sqrt{(h-1)(h-\gamma)}$. From here, we readily find the action-angle variables $(\phi_{0},I)$ and write the initial conditions in terms of them as
\begin{equation}
Q_{0}=\sqrt{\frac{2I\sqrt{h-\gamma}}{\sqrt{h-1}}}\,\sin\phi_{0},\,\,\,\,\,\,\,P_{0}=\sqrt{\frac{2I\sqrt{h-1}}{\sqrt{h-\gamma}}}\,\cos\phi_{0}.\label{initc}
\end{equation}
We substitute~(\ref{initc}) in~(\ref{solLipkin}) and use~(\ref{classDef}) with the deformation functions (\ref{defLipkin}) to find the classical metric. The resulting metric components are
\begin{subequations}
\label{classMetricSym}
\begin{align}
g_{11} & =\frac{I^{2}}{32}\left[\frac{1-\gamma}{(h-1)(h-\gamma)}\right]^{2}, \\
g_{12} & =\frac{I^{2}(1-\gamma)}{32(h-1)(h-\gamma)^{2}}, \\
g_{22} & =\frac{I^{2}}{32(h-\gamma)^{2}}.
\end{align}
\end{subequations}
We see that at the QPT, $h=1$, the components $g_{11}$ and $g_{12}$ diverge. Nevertheless, we can also note that the determinant of the classical metric is zero.

Now, to compute the QMT in the thermodynamic limit, we use the Hamiltonian (\ref{HLipkin}) which leads to the quantum deformation operators
\begin{subequations}
\label{QdefSym}
\begin{align}
\hat{{\cal O}}_{1} & =\frac{\partial\hat{H}}{\partial h}=-2\hat{J}_{z}, \\
\hat{{\cal O}}_{2} & =\frac{\partial\hat{H}}{\partial\gamma}=-\frac{\hat{J}_{y}^{2}}{j}.
\end{align}
\end{subequations}
Recall that when $j\rightarrow\infty$, the truncated Holstein-Primakoff transformation allows us to cast the angular momentum operators in the $(\hat{Q},\hat{P})$ representation, as suggested by Eq.~(\ref{QyP}). Thus, the quantum deformation operators read as
\begin{subequations}
\label{defsym}
\begin{align}
\hat{{\cal O}}_{1} & =\hat{P}^{2}+\hat{Q}^{2}-2j-1, \\
\hat{{\cal O}}_{2} & =-\hat{P}^{2}.
\end{align}
\end{subequations}
We express them in terms of creation and annihilation operators and time, and read off the spectrum from the effective quadratic Hamiltonian~(\ref{quantQP}). This information is then substituted into Eq.~(\ref{qmetric}), which yields the following metric components of the QMT:
\begin{subequations}
\label{QuantMetricSym}
\begin{align}
g_{11}^{(0)} & =\frac{1}{32}\left[\frac{1-\gamma}{(h-1)(h-\gamma)}\right]^{2}, \\
g_{12}^{(0)} & =\frac{1-\gamma}{32(h-1)(h-\gamma)^{2}}, \\
g_{22}^{(0)} & =\frac{1}{32(h-\gamma)^{2}}.
\end{align}
\end{subequations}
This QMT has already been obtained in Ref.~\cite{Sarkar2012} by using another method. Hence, we corroborate it with our approach. We also see that its determinant is zero, which was noted in the same reference and implies that information geometry is ill defined in the symmetric phase of this model. Comparing Eqs.~(\ref{classMetricSym}) and~(\ref{QuantMetricSym}), it is easy to see that the classical and quantum results have exactly the same parameter dependence and that the classical metric reproduces the singularities of the QMT. Moreover, both metrics match perfectly if the identification $I^{2}=1$ is made. This is a remarkable result, because through a classical procedure, like the parametrization of the canonical coordinates $(Q,P)$ in terms of action-angle variables [see Eq.~(\ref{initc})] and the classical torus average~(\ref{classAv}), we have been able to obtain the quantum result.

\subsubsection{\textbf{Metrics for the broken phase}}

For the broken phase, the classical Hamiltonian is obtained by taking the expectation value of Eq.~(\ref{Hbroken}) in spin coherent states $|z\rangle$, which leads to
\begin{align}
H_{cl}^{\prime}= & -j(1+h^{2})+(1-\gamma)P^{2}+(1-h^{2})Q^{2} \nonumber \\
& +\frac{h}{\sqrt{j}}\sqrt{1-h^{2}}Q(P^{2}+Q^{2})\sqrt{1-\frac{P^{2}+Q^{2}}{4j}} \nonumber \\
& +\frac{1}{4j}(P^{2}\!+\!Q^{2})\left[\gamma P^{2}\!+\!h^{2}Q^{2}\!-(1\!-\!h^{2})(P^{2}\!+\!Q^{2})\right].\label{Hclassrot}
\end{align}
The quadratic approximation of this Hamiltonian is given by
\begin{equation}
H_{cl}^{\prime}\simeq-j(1+h^{2})+(1-\gamma)P^{2}+(1-h^{2})Q^{2}.
\end{equation}
In this case, the evaluation of $H_{cl}^{\prime}$ at $Q=P=0$ yields the broken phase ground state $E_{g}=-j(1+h^{2})$. The deformation functions are the same as those of~(\ref{defLipkin}); however, they must be expressed in terms of rotated quantities, in which case they take the following form:
\begin{subequations}
\begin{align}
{\cal O}_{1}^{\prime} & =2\sqrt{1-h^{2}}J_{x}^{\prime}-2hJ_{z}^{\prime}, \\
{\cal O}_{2}^{\prime} & =-\frac{J_{y}^{\prime2}}{j}.
\end{align}
\end{subequations}
From here, we move to the position-momentum representation through the transformation~(\ref{transfQPLipkin}) and find that the deformation functions are
\begin{subequations}
\label{defbroken}
\begin{align}
{\cal O}_{1}^{\prime} & =h(P^{2}+Q^{2})+2\sqrt{j(1-h^{2})}Q-2hj-h, \\
{\cal O}_{2}^{\prime} & =-P^{2}.
\end{align}
\end{subequations}
Then, we use~(\ref{solLipkin}) and~(\ref{initc}), and substitute~(\ref{defbroken}) into~(\ref{classDef}), which yields the components of the classical metric
\begin{subequations}
\label{classMetricBroken}
\begin{align}
g_{11} & =\frac{jI}{\sqrt{(1-h^{2})(1-\gamma)}}+\frac{I^{2}}{32}\left[\frac{h(h^{2}-\gamma)}{(1-h^{2})(1-\gamma)}\right]^{2}, \\
g_{12} & =\frac{I^{2}h\left(h^{2}-\gamma\right)}{32\left(1-h^{2}\right)(1-\gamma)^{2}}, \\
g_{22} & =\frac{I^{2}}{32(1-\gamma)^{2}}.
\end{align}
\end{subequations}
As for the QMT in the broken phase, we use the deformation operators~(\ref{QdefSym}), rewrite them in terms of the rotated angular momenta~(\ref{rot}), and use the truncated Holstein-Primakoff transformation~(\ref{QyP}) to compute the metric. The resulting QMT is
\begin{subequations}
\label{metricbroken}
\begin{align}
g_{11}^{(0)} & =\frac{j}{2\sqrt{(1-h^{2})(1-\gamma)}}+\frac{1}{32}\left[\frac{h(h^{2}-\gamma)}{(1-h^{2})(1-\gamma)}\right]^{2}, \\
g_{12}^{(0)} & =\frac{h(h^{2}-\gamma)}{32(1-h^{2})(1-\gamma)^{2}}, \\
g_{22}^{(0)} & =\frac{1}{32(1-\gamma)^{2}}.
\end{align}
\end{subequations}
It is remarkable that both classical and quantum metrics have the exactly same parameter structure and perfectly match using the identifications of the action variables. In contrast to the symmetric phase, both metrics are now invertible and have the determinants
\begin{subequations}
\begin{align}
\det g&=\frac{I^{3}}{32\sqrt{(1-h^{2})(1-\gamma)^{5}}}, \\
\det g^{(0)}&=\frac{j}{64\sqrt{(1-h^{2})(1-\gamma)^{5}}},
\end{align}
\end{subequations}
which at the critical point $h=1$ diverge. This is a further result since Ref.~\cite{Sarkar2012} did not analyze the broken phase of the model. Thus, we have found that the broken phase has a well-defined metric structure that allows a further geometric characterization with the aid of the scalar curvature. Furthermore, we see once more that the classical metric contains the whole information that can be extracted from the QMT, with the advantage that it is simpler to compute. The scalar curvature for either of these two metrics can be computed with Eq.~(\ref{defscalar}), which yields
\begin{equation}
R=-4+\frac{7h^{4}-(9\gamma-2)h^{2}-4(1-\gamma)}{j\sqrt{(1-h^{2})(1-\gamma)^{3}}}.\label{RLip}
\end{equation}
From this expression, we observe that for large values of $j$ (as is expected in the thermodynamic limit), the scalar curvature practically takes on the constant value $-4$, and that it diverges at $h=1$, which indicates the presence of the QPT. It is interesting to observe that the metric's singularity is independent of the coordinate system for this phase since it also appears in the scalar curvature which is a geometric invariant~\cite{Villani2008}.

\subsubsection{\textbf{Numerical analysis for finite $j$}}

We now want to address the effects of having a finite $j$ directly and without resorting to any approximations. This will help us delve into the nature of the singularities of the QMT and the scalar curvature, and deduce whether they are effects of the truncated Holstein-Primakoff transformation or are intrinsic to the system. It is worth mentioning that, to our knowledge, the numerical computation of the scalar curvature for the LMG has not been carried out previously.

The Hamiltonian~(\ref{HLipkin}) can be numerically diagonalized, which will give us in turn the differences between the classical metric [Eqs.~(\ref{classMetricSym}) and~(\ref{classMetricBroken})] with the QMT for a given value of $j$. We obtain the numerical results by employing the so-called perturbative form of the QMT, which reads as
\begin{equation}
g_{ij}^{(0)}(x)=\sum_{n\neq0}\frac{\langle0|\hat{{\cal O}}_{i}|n\rangle\langle n|\hat{{\cal O}}_{j}|0\rangle}{(E_{n}-E_{0})^{2}}.\label{QMTpert}
\end{equation}
The evaluation of this formula requires time-independent deformation operators, as opposed to Eq.~(\ref{qmetric}), but at the cost of summing over all the elements of the eigenspace of $\hat{H}$ (for details, see~\cite{GuReview}).

\begin{figure}[ht]
\includegraphics[width=\columnwidth]{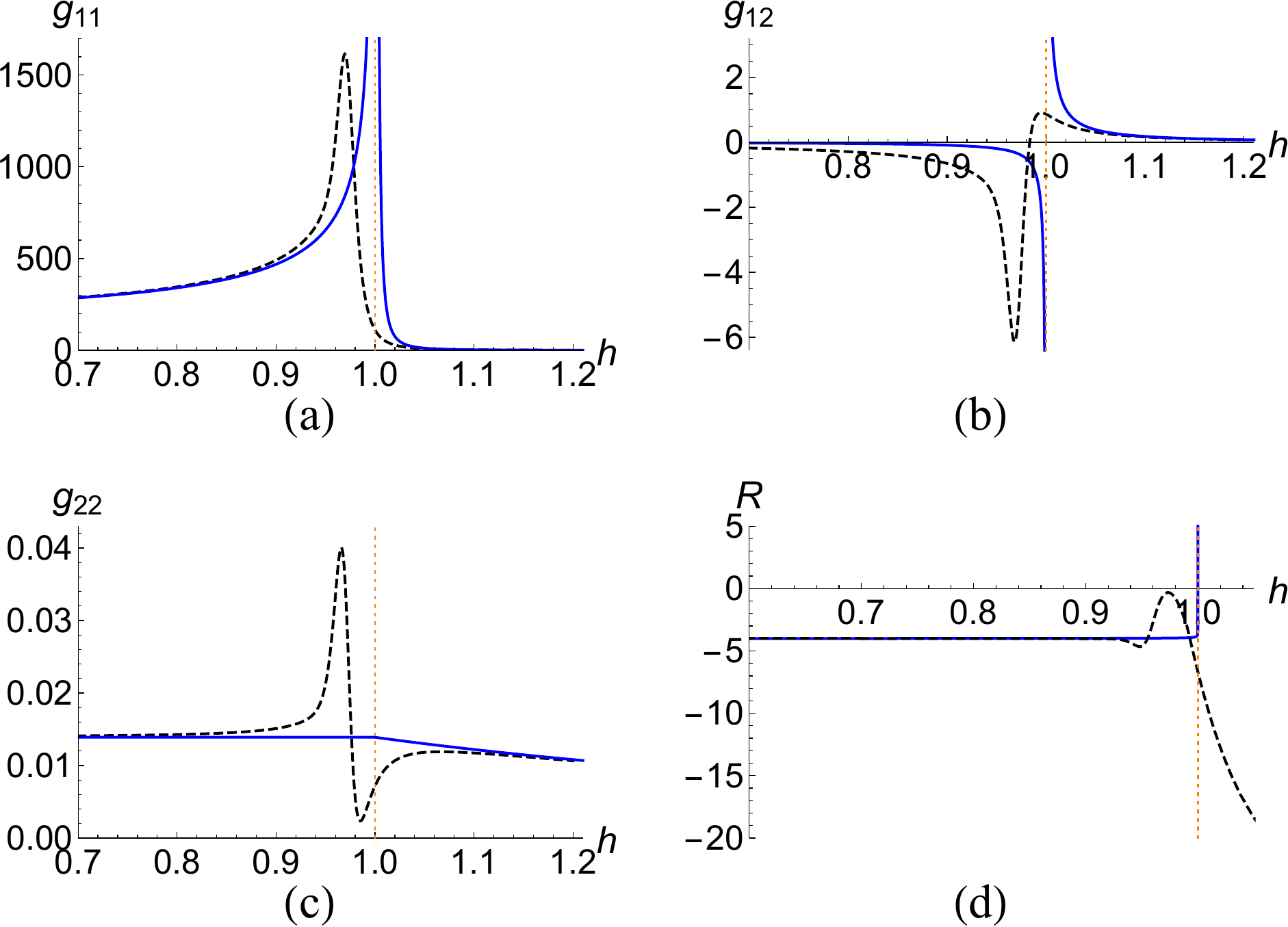}
\caption{Metric components and scalar curvature of the classical (or quantum) metric in the thermodynamic limit (solid blue) and the exact QMT for $j=500$ (dashed black) as a function of $h$ when $\gamma=-0.5$. In the thermodynamic limit, the $g_{11}$ and $g_{12}$ components and the scalar curvature diverge at the phase transition (dotted orange).}
\label{FigMetric_vs}
\end{figure}

We begin by comparing, in Figs.~\hyperref[FigMetric_vs]{4(a)--4(c)}, the classical (or quantum) metric in the thermodynamic limit\footnote{Recall that the classical and quantum metrics yield the same result in the thermodynamic limit.} with the exact QMT [obtained through~(\ref{QMTpert})] for $j=500$ and $\gamma=-0.5$. We see that the agreement between them is acceptable as long as we are not close to the QPT where the Holstein-Primakoff approximation fails~\cite{Hirsch2013}. At the transition, the analytic metric components $g_{11}$ and $g_{12}$ show a divergence that is not present in their finite $j$ counterparts coming from Eq.~(\ref{QMTpert}). For the scalar curvature, we see in Fig.~\hyperref[FigMetric_vs]{4(d)} that the analytic plot has a divergence at $h=1$, and that it does not exist in the region $h>1$, which was expected from Eq.~(\ref{RLip}).

\begin{figure}[ht]
\includegraphics[width=\columnwidth]{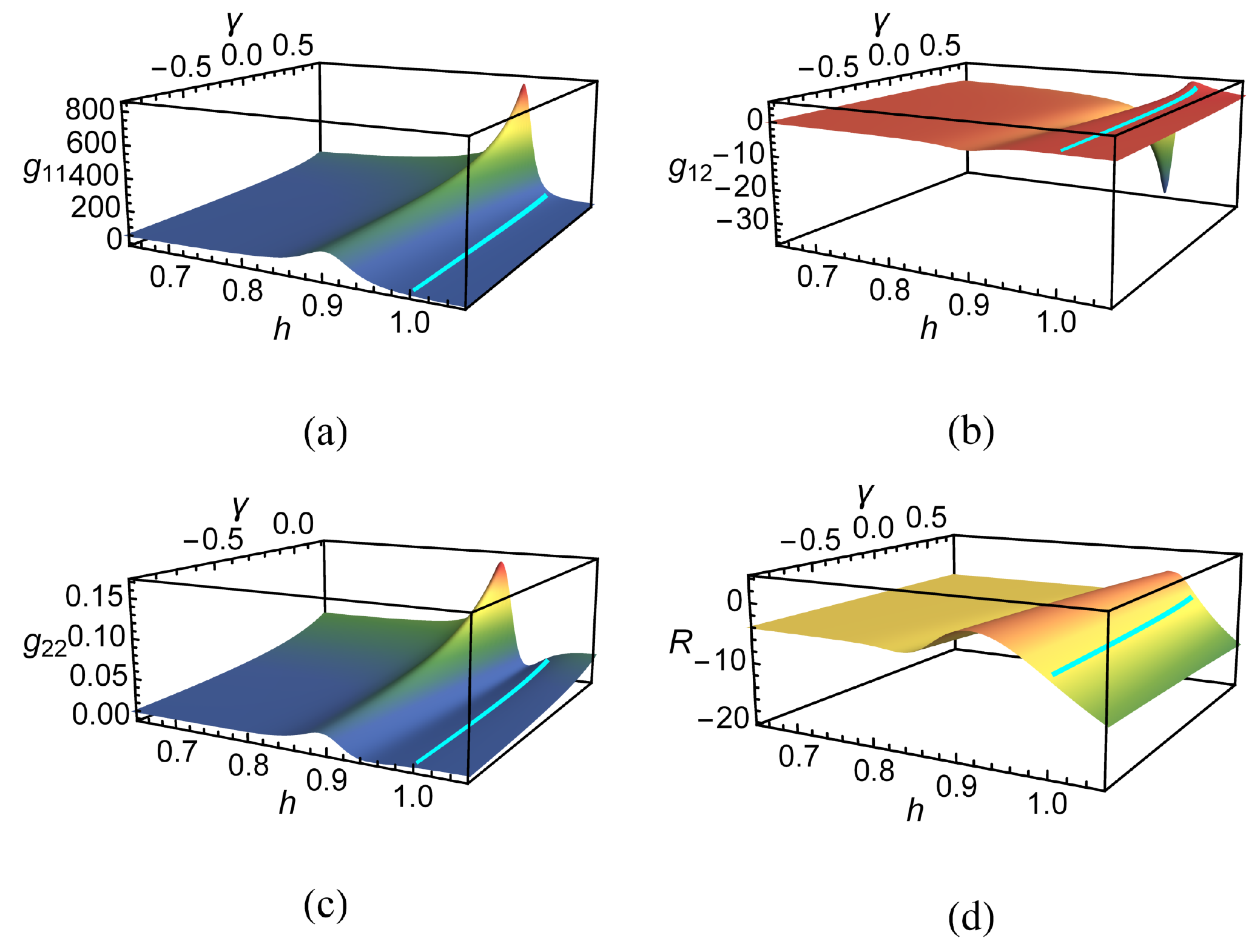}
\caption{Metric components and scalar curvature for $j=100$. The plots clearly show the presence of the QPT precursors. The critical line $h=1$ is shown in cyan.}
\label{FigMetric_3D}
\end{figure}

\begin{figure}[ht]
\includegraphics[width=\columnwidth]{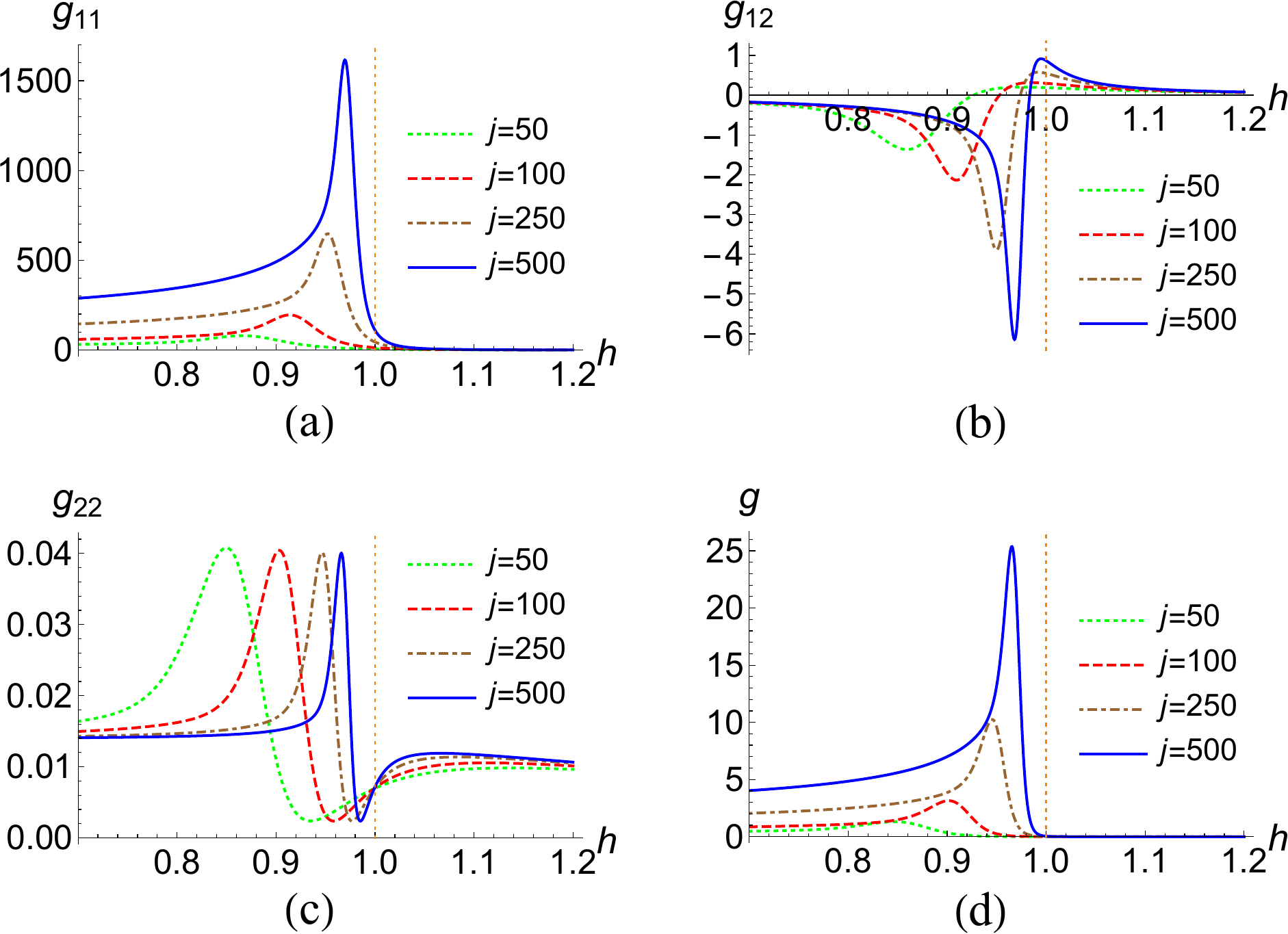}
\caption{QMT components and determinant for different values of $j$ when $\gamma=-0.5$. The peaks in the metric components and in the determinant become narrower as $j$ increases.}
\label{FigMetric_js}
\end{figure}

In Fig.~\ref{FigMetric_3D}, we show the three components of the QMT for $j=100$ and the scalar curvature obtained through numerical differentiation over a mesh in the parameter space. We see the appearance of peaks near $h=1$, which we identify as the precursors of the QPT. This is most clearly seen in Fig.~\ref{FigMetric_js}, where we show the QMT and its determinant for a fixed $\gamma$ and different values of $j$. We notice that the peaks of both the metric components and its determinant become narrower and get closer to $h=1$ as $j$ increases, which corroborates their identification as the precursors of the QPT. Actually, this suggests that for large values of $j$ and $h=1$, the exact QMT components $g_{11}$ and $g_{12}$ will have a singularity. On the other hand, $g_{22}$ does not seem to have such a good agreement with its analytic counterpart near the QPT; this is because $g_{22}$ [see Eqs.~(\ref{classMetricSym}) and~(\ref{classMetricBroken})] is not sensitive to the critical value $h=1$, unlike the other components.

\begin{figure}[ht]
\includegraphics[width=\columnwidth]{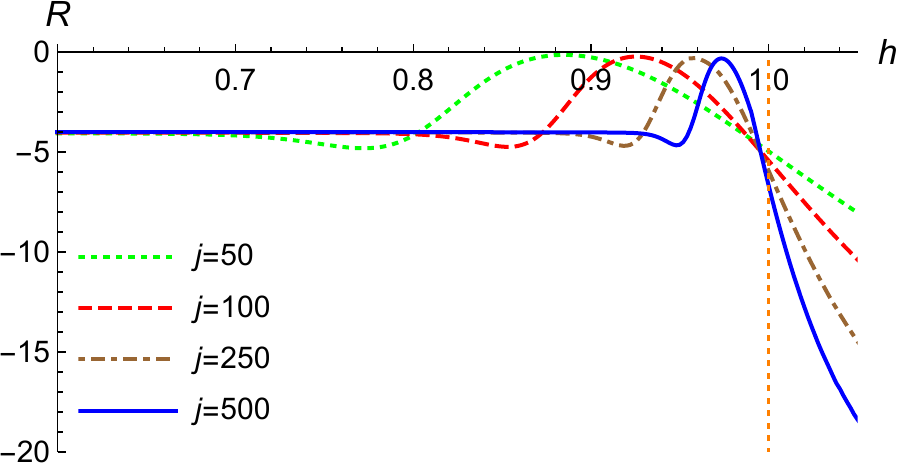}
\caption{Scalar curvature for different values of $j$ when $\gamma=-0.5$. It is seen that the two peaks (one negative and one positive) approach each other and that the slope of $R$ around $h=1$ gets steeper as $j$ increases.}
\label{FigR_js}
\end{figure}

\begin{figure}[ht]
\includegraphics[width=\columnwidth]{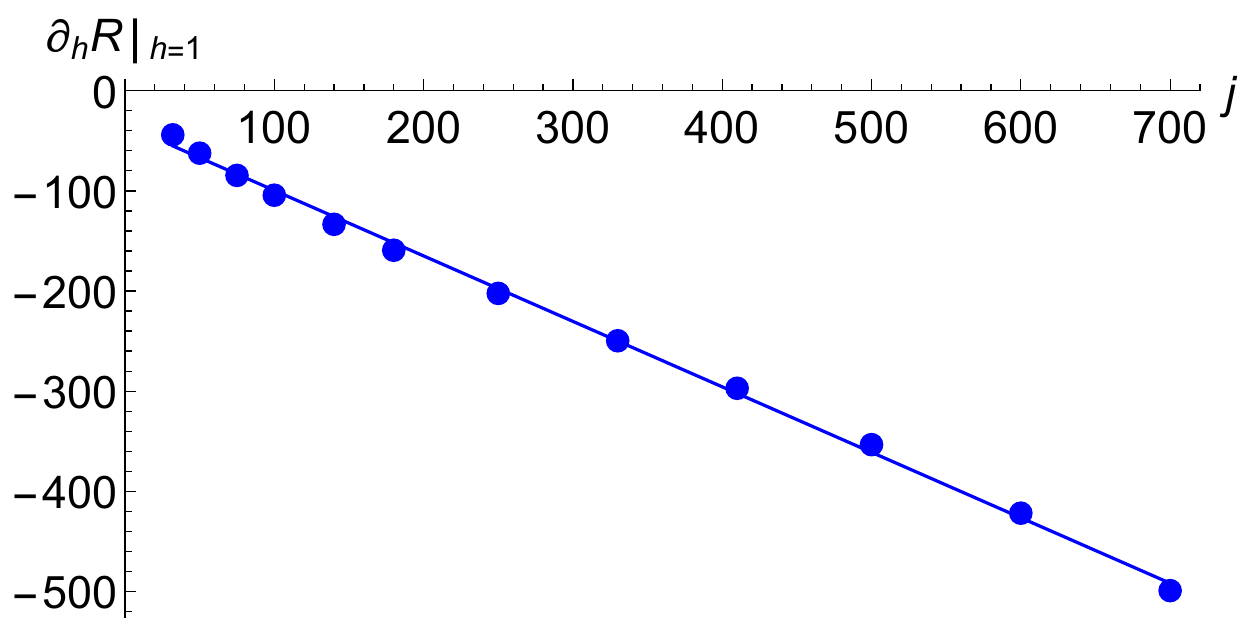}
\caption{Derivative of the scalar curvature with respect to $h$ evaluated at $h=1$ for different values of $j$ when $\gamma=-0.5$. The plot clearly shows the tendency of the slope to higher negative values.}
\label{FigRder_js}
\end{figure}

Now, Fig.~\hyperref[FigMetric_js]{6(d)} helps us understand the behavior of the scalar curvature (see Fig.~\ref{FigR_js}) when $h>1$. The determinant falls rapidly to near-zero values just before the separatrix and maintains these values when $h>1$. That is why we see in Figs.~\ref{FigR_js} and~\ref{FigRder_js} that the slope of the descending curve for $h>1$ gets steeper as $j$ increases. Thus, we expect that for large values of $j$ and $h=1$, the scalar curvature will be an almost vertical line that falls off to large negative values. The exhibited dissimilar behavior between the analytic and numerical results is a consequence of the failure of the Holstein-Primakoff truncation at $h=1$, which causes the scalar curvature to diverge, in contrast to the smooth numerical curve.

In Fig.~\ref{FigRder_js}, we analyze in more detail the behavior of the derivative of $R$ with respect to $h$ evaluated at $h=1$ as a function of $j$ for $\gamma=-0.5$. The function that fits the points is
\begin{equation}
\frac{\partial R}{\partial h}\bigg{|}_{h=1}=-34.2-0.65 j.
\end{equation}
This equation clearly shows that the derivative of $R$ goes to $-\infty$ as $j\to\infty$, confirming the behavior displayed in Fig.~\ref{FigR_js}. Therefore, we infer that in the thermodynamic limit, the discontinuity of the scalar curvature (which is $-4$ in the symmetric phase and does not exist in the broken phase) is the cause of the singularity in the derivative at the QPT. In another work~\cite{Vergara2021PRB}, we explore a modified LMG model that has an invertible metric in both phases and makes this point clearer, showing without doubt that the scalar curvature is discontinuous at the QPT in the thermodynamic limit, which causes the divergence in its derivative there.

Having studied the slope of $R$ at the QPT, we analyze in the next section the behavior of the QMT and its scalar curvature in terms of the maxima and minima that they display.

\subsubsection{\textbf{Peak analysis}}

\begin{figure}[ht]
\includegraphics[width=\columnwidth]{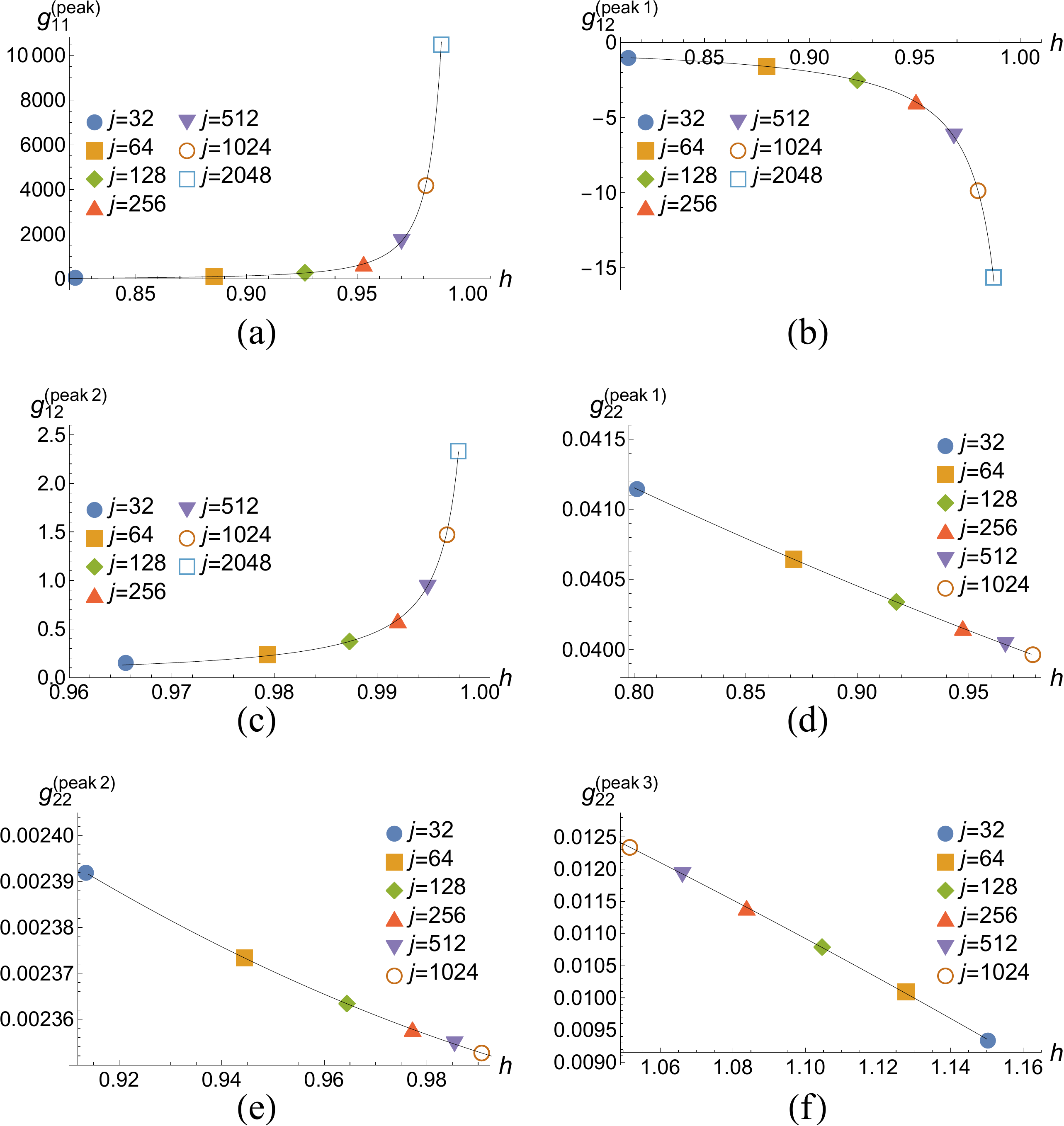}
\caption{Peaks of the QMT components as functions of $h$ when $\gamma=-0.5$. The value of $j$ is indicated for each point. The $g_{11}$ component has only one peak, $g_{12}$ has two peaks, and $g_{22}$ has three peaks.}
\label{FigPeaks_h}
\end{figure}

To better understand the behavior of numerical QMT and its scalar curvature for finite $j$, we close this section with an analysis of their peaks. In Fig.~\ref{FigPeaks_h}, we plot the height of the peaks of the exact QMT components as a function of $h$ while we fix $\gamma=-0.5$. The curves that interpolate the points have the following form:

(i) $g_{11}$ peak:
\begin{equation}	\label{g11peak}
g_{11}^{\rm(peak)}=-22.5317+\frac{1.5333}{(h-1)^2}.
\end{equation}

(ii) $g_{12}$ peaks:
\begin{subequations}
	\label{g12peaks}
\begin{align}
g_{12}^{\rm(peak\,1)}&=0.0608+\frac{0.1990}{h-1}, \\
g_{12}^{\rm(peak\,2)}&=-0.0103-\frac{0.0048}{h-1}.
\end{align}
\end{subequations}

(iii) $g_{22}$ peaks:
\begin{subequations}
\begin{align}
g_{22}^{\rm(peak\,1)}&=0.0498-0.0142h+0.0042h^2, \\
g_{22}^{\rm(peak\,2)}&=0.0046-0.0042h+0.0019h^2, \\
g_{22}^{\rm(peak\,3)}&=0.0207-0.0125h-0.0194h^2.
\end{align}
\end{subequations}
From the functions~(\ref{g11peak}) and ~(\ref{g12peaks}), it is clear that in the limits $h=1$ and $j\rightarrow\infty$ the QMT components $g_{11}$ and $g_{12}$ exhibit a divergent behavior, which is in accordance with the corresponding classical (or quantum) metric components in the thermodynamic limit.

\begin{figure}[ht]
\includegraphics[width=\columnwidth]{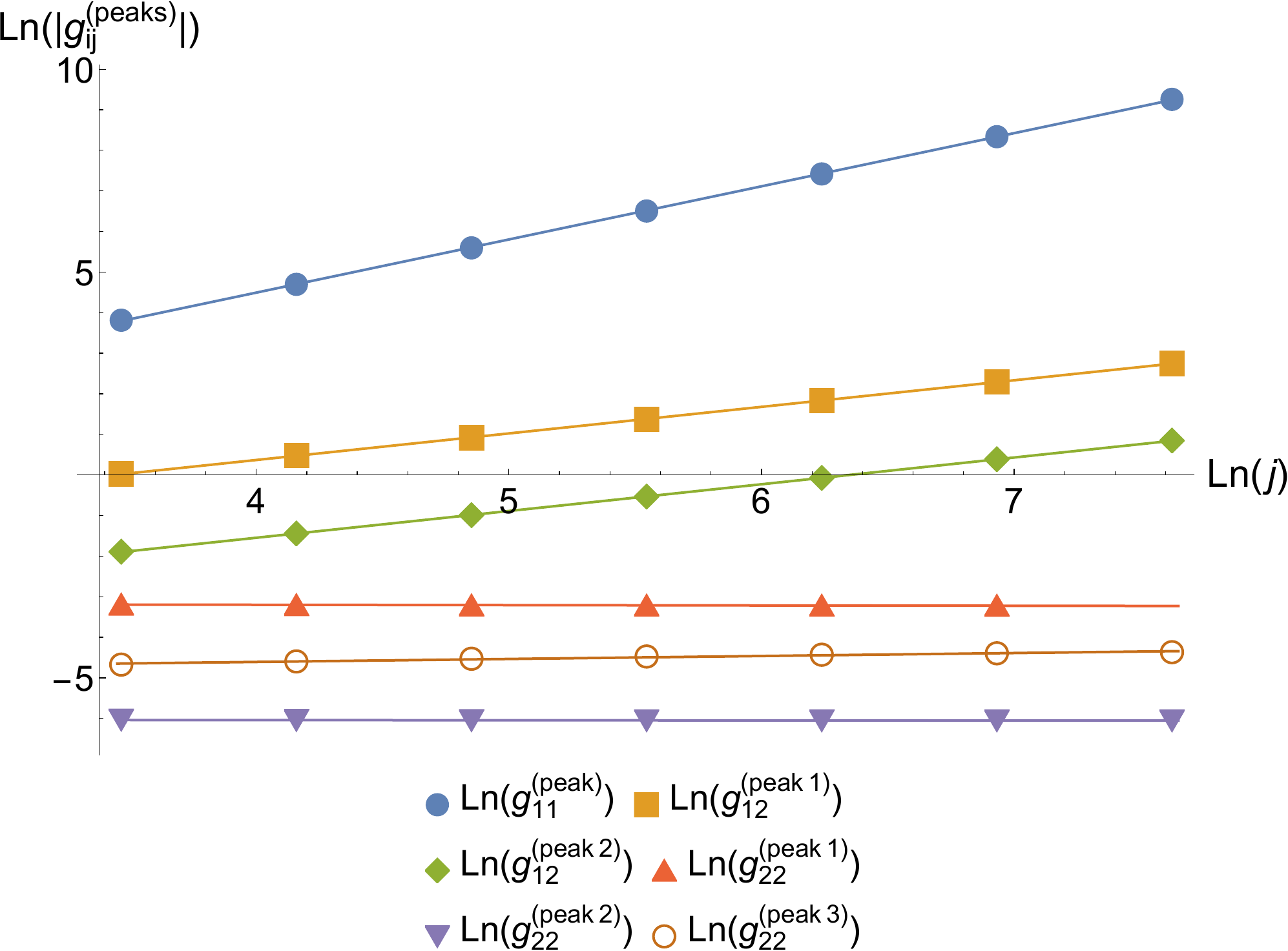}
\caption{Peaks of the QMT components as functions of $j$ when $\gamma=-0.5$. The axes are presented in logarithmic scale.}
\label{FigPeaks_j}
\end{figure}

In Fig.~\ref{FigPeaks_j}, we plot the height of the peaks as a function of $j$ for $\gamma=-0.5$. A linear relation between these quantities is evident when using a log-log scale. The function we use to fit the points is
\begin{equation}
\ln(g_{ij}^{\rm(peak)})=m \ln(j)+n,\label{logFunction}
\end{equation}
where the parameters $m$ and $n$ are shown in Table~\ref{Table} for every peak. Notice that we reproduce the value $m\approx 1.3$ for the metric component $g_{11}$ that was obtained in~\cite{GuPRE,Vidal2004,Vidal2005}. In addition to this, we analyze the other components, finding that the sum of the $m$ values for the two peaks of $g_{12}$ is $1.3142$, which a similar result to that of the $g_{11}$ component. This is because $g_{12}$ has mixed information about the parameters $h$ and $\gamma$. Accordingly, the $m$ values of $g_{22}$ do not relate with those of the other components.

\begin{table}[ht]
\caption{\label{Table} Values of $m$ and $n$ in Eq.~(\ref{logFunction}) for $\gamma=-0.5$. Notice that the two peaks of $g_{12}$ have the property that their values of $m$ sum up to 1.3142, which is close to the $m$ value of $g_{11}$ peak.}
\begin{ruledtabular}
\begin{tabular}{ccc}
Peak & $m$ & $n$ \\
\hline
$g_{11}^{(\rm peak)}$ & 1.3103 & -0.7480 \\
$g_{12}^{(\rm peak\,1)}$ & 0.6549 & -2.2513 \\
$g_{12}^{(\rm peak\,2)}$ & 0.6593 & -4.1853 \\
$g_{22}^{(\rm peak\,1)}$ & -0.0081 & -3.1676 \\
$g_{22}^{(\rm peak\,2)}$ & -0.0037 & -6.0269 \\
$g_{22}^{(\rm peak\,3)}$ & 0.0731 & -4.9007 \\
\end{tabular}
\end{ruledtabular}
\end{table}

\begin{figure}[ht]
\includegraphics[width=\columnwidth]{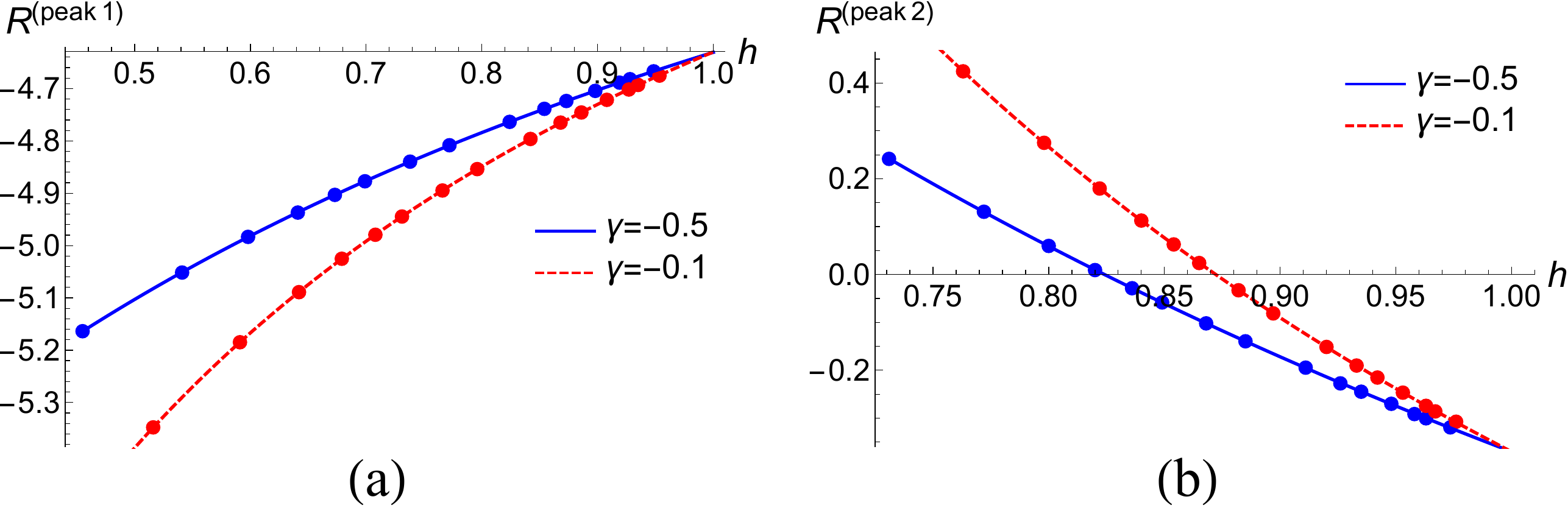}
\caption{Peaks of the scalar curvature as functions of $h$ for $\gamma=-0.5$ and $-0.1$. The values of $j=$12,16,20,24,28,32,40,50,75,100,125,\\175,250,300,500 were considered. In both plots, $j$ grows with $h$.}
\label{FigPeaksR_h}
\end{figure}

Now, we study the scalar curvature for two representative values of $\gamma$. This will allow us to characterize $R$ for finite $j$ and infer its behavior in the $j\rightarrow\infty$ limit. We first analyze the behavior of the extrema as a function of $h$. In Fig.~\ref{FigPeaksR_h}, we plot the two peaks of $R$ for $\gamma=-0.5$ and $-0.1$. One of the peaks is a local minimum (peak 1) and the other is a maximum (peak 2). The functions that interpolate the points of the minima are
\begin{subequations}
\label{Rpeak1h}
\begin{align}
R^{(\rm peak\,1)}(h,\gamma=-0.5)&=-3.407-\frac{2.197}{h+0.795}, \\
R^{(\rm peak\,1)}(h,\gamma=-0.1)&=-3.443-\frac{1.525}{h+0.285},
\end{align}
\end{subequations}
whereas for the maxima they are
\begin{subequations}
\label{Rpeak2h}
\begin{align}
R^{(\rm peak\,2)}(h,\gamma=-0.5)&=-2.972+\frac{3.674}{h+0.412}, \\
R^{(\rm peak\,2)}(h,\gamma=-0.1)&=-2.685+\frac{2.145}{h-0.073}.
\end{align}
\end{subequations}
At $h=1$, where the QPT appears in the thermodynamic limit, the first peak of $R$ takes the value of $-4.631$ when $\gamma=-0.5$ and $-4.630$ when $\gamma=-0.1$, whereas the second peak goes to $-0.370$ when $\gamma=-0.5$ and to $-0.371$ when $\gamma=-0.1$. The value of the first peak in both cases is very close to the one predicted by the truncated Holstein-Primakoff approximation. Also, notice that at $h=1$, the minima for the two different values of $\gamma$ are similar, which is also the case for the maxima. This is seen in Fig.~\ref{FigPeaksR_h}, where both lines intersect.

\begin{figure}[ht]
\includegraphics[width=\columnwidth]{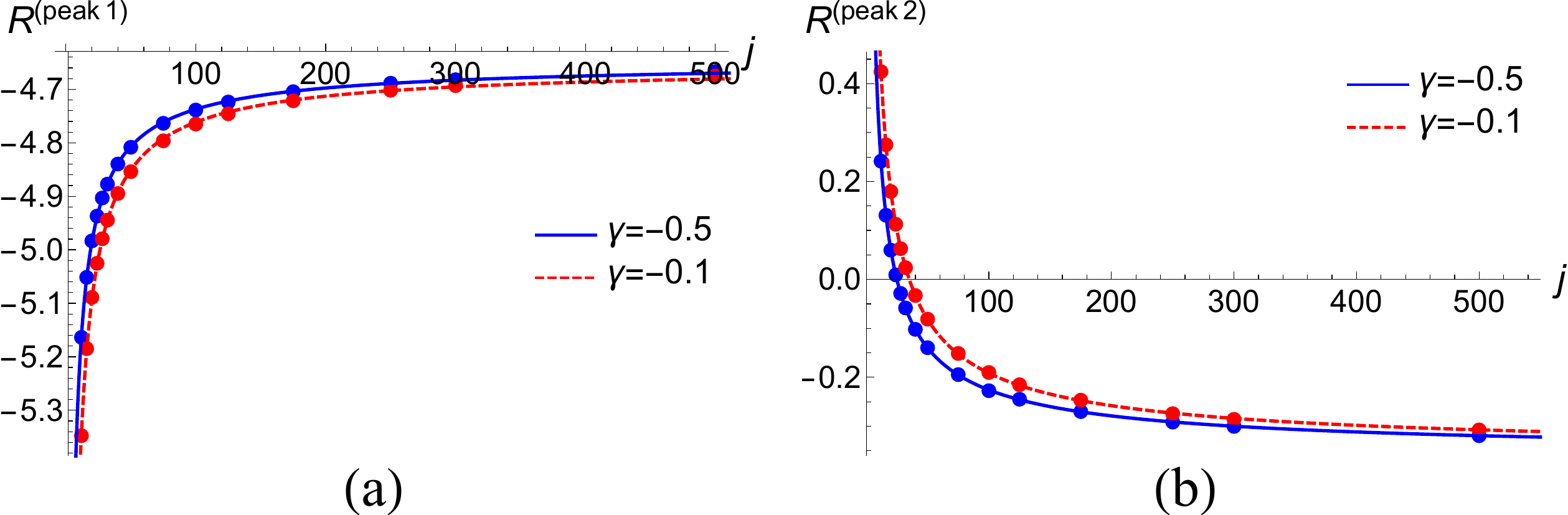}
\caption{Peaks of the scalar curvature as functions of $j$ for $\gamma=-0.5$ and $-0.1$. The minimum is shown in (a) and the maximum is shown in (b).}
\label{FigPeaksR_j}
\end{figure}

If we now analyze the value of the peaks as a function of $j$, we can fit the data for the minima as follows:
\begin{subequations}
\label{Rpeak1j}
\begin{align}
R^{(\rm peak\,1)}(j,\gamma=-0.5)&=-4.645-\frac{3.882}{j^{0.812}}, \\
R^{(\rm peak\,1)}(j,\gamma=-0.1)&=-4.655-\frac{6.100}{j^{0.879}}.
\end{align}
\end{subequations}
For the maxima, the functions take the form
\begin{subequations}
\label{Rpeak2j}
\begin{align}
R^{(\rm peak\,2)}(j,\gamma=-0.5)&=-0.365+\frac{3.408}{j^{0.695}}, \\
R^{(\rm peak\,2)}(j,\gamma=-0.1)&=-0.360+\frac{4.749}{j^{0.726}}.
\end{align}
\end{subequations}
We show in Fig.~\ref{FigPeaksR_j} the behavior of the two peaks of $R$ as a function of $j$ when $\gamma=-0.5$ and $-0.1$. Two main features are seen. First, the minimum (peak 1) grows as $j$ increases and the maximum (peak 2) decreases with $j$. Second, the functions~(\ref{Rpeak1j}) reveal that the minimum reaches a value around $-4.6$ when $j\rightarrow\infty$, whereas the maximum~(\ref{Rpeak2j}) goes to $-0.36$. This means that both peaks persist in the thermodynamic limit, which is also consistent with Eqs.~(\ref{Rpeak1h}) and~(\ref{Rpeak2h}), since the limits $h=1$ and $j\rightarrow\infty$ in both sets of equations predict the same results. Therefore, although the scalar curvature is smooth near the QPT, its peaks serve as precursors of the phase transition for finite $j$. Furthermore, in contrast to the QMT which is singular at the QPT in the thermodynamic limit, the scalar curvature is discontinuous there.

A final comment can be made regarding the change of sign in the maximum of $R$. Solving the condition $R^{(\rm peak\,2)}=0$ yields the value of $j$ for which the maximum is zero. For $\gamma=-0.5$ this value is $j=25$, while for $\gamma=-0.1$ the result is $j=35$. This indicates a local change in the geometry of the parameter space, i.e., a change in curvature from spherical type to hyperbolic type, although we do not attribute any special interpretation to those values of $j$.

\section{Conclusions\label{sec:Conclusions}}

We studied in this paper the classical analog of the QMT and its scalar curvature for the Dicke and LMG models. Our results confirm that the classical torus average reproduces all or almost all the parameters' structure of the QMT. In the Dicke model, we considered the thermodynamic limit under the truncated Holstein-Primakoff approximation and showed that the classical metric and its scalar curvature have similar behavior as their quantum counterparts near the QPT. The classical metric was obtained in this case, and the QMT was corroborated under our time-dependent deformation functions approach. Also, the resonance condition was analyzed, and a divergence in the scalar curvature was found, as opposed to the nonresonant case; here, too, both the classical and quantum scalars diverge in the same manner. In the LMG model, our analysis consisted of two main parts. First, we calculated the classical metric and the QMT in the thermodynamic limit and showed that they are identical, modulo a quantization rule for action variables. Second, we computed the QMT for finite $j$ and compared it with the analytic counterpart, resulting in a remarkable agreement except near the QPT, where the truncated Holstein-Primakoff transformation is not valid anymore. In Ref.~\cite{GuPRE}, the authors found the fidelity susceptibility related to the parameter $h$. However, we contributed by  exploring the parameter $\gamma$ through the other two components of the QMT and closed the geometric study with the analytic and numerical computation of the scalar curvature, which had not been performed for this model. The numerical results showed that both the QMT and the scalar curvature are smooth for finite $j$. The numerical analysis was further expanded by describing the peaks of the metric components and the scalar curvature near the critical region, which provided us with one last crucial result: the maxima's and minima's dependence on $h$ and $j$. This analysis showed that the peaks of the QMT as well as of the scalar curvature are the QPT precursors for finite $j$. Furthermore, the extrapolation to $j\rightarrow\infty$ showed that the QMT is singular at the QPT and that {\it the scalar curvature is discontinuous there}.  For the Dicke model, the classical metric's behavior is the same as the quantum one near the QPT, and the scalar curvature has the same value there regardless of the presence of an anomaly. Remarkably, the effect of the anomaly is not that relevant near the QPT. This is worth pointing out since through a renormalization procedure in the critical region, the thermodynamic limit results can provide information regarding the scaling properties for finite $j$ and help classify the models in universality classes~\cite{Hirsch2013}. All these features support our claim that the classical metric can give a preliminary or even a complete idea of the quantum result.

The use of the classical metric opens the way to extend our present work. For instance, we could study the quantum Dicke Hamiltonian for finite $j$~(\ref{Dickeoriginal}), and compare the results with those obtained here. Furthermore, we could go deeper into the classical setting and explore the full mean-field Dicke Hamiltonian constructed with coherent states, which is far more involved and may even contain chaotic dynamics (see Refs. \cite{Bastarrachea2015,Hirsch2019}). A first way to proceed would be to carry out a classical perturbative analysis as in Ref.~\cite{GonzalezPRE} to find corrections to the quadratic approximation in the thermodynamic limit and compare the results with their quantum counterpart~\cite{AlvarezGenerating}. Besides, the chaotic region could be approached by studying the adiabatic gauge potential (which is deeply related to the QMT) since it has been found recently that it serves as a sensitive measure of quantum chaos~\cite{Polkovnikov2020}. In this regard, the classical metric may also be of interest to study quantum scarring in the Dicke model~\cite{Hirsch2021}.

A different perspective in the study of quantum systems can also be addressed. We could use the Wigner function formulation of the QMT~\cite{Gonzalez2020} to study a variety of many-body systems and see whether it provides a deeper insight not only into the QPTs, which refer to the ground state, but also into excited-state quantum phase transitions (ESQPTs)~\cite{Caprio2008}. Additionally, the Wigner function formalism may shed some light on the semiclassical approximation and help clarify the anomalies' role that accounts for the difference between the classical and quantum results. Furthermore, the full mean-field LMG Hamiltonian~(\ref{LipkinClassQP}) could be addressed and establish the similarities and differences with the truncated results of the Holstein-Primakoff transformation. Of course, the exploration of the classical metric, the QMT, and their scalar curvature for plenty other models from quantum optics and condensed matter can be carried out, and hopefully, the results will provide us with a geometric picture that will allow a deeper understanding of the dynamics of these systems. Moreover, it will be interesting to compute quantum, and classical metrics in the case of Weyl and Dirac semimetals \cite{RevModPhys.90.015001} and observe if the classical metric can detect the chiral anomalies.

\acknowledgments

D. Guti\'errez-Ruiz is supported by a CONACyT Ph.D. Scholarship No. 332577. D. Gonzalez was partially supported by a DGAPA-UNAM postdoctoral fellowship and by Consejo Nacional de Ciencia y Tecnolog\'ia (CONACyT), M\'exico, Grant No. A1-S-7701. This work was partially supported by DGAPA-PAPIIT Grant No. IN103919.

\end{document}